\documentclass[preprint,12pt]{elsarticle}

\usepackage{graphicx} % Required for inserting images
\usepackage{amsmath}
\usepackage{amsfonts}
\usepackage{natbib}
\usepackage{booktabs}
\usepackage{multirow}
\usepackage{algorithm}
\usepackage{algpseudocode}
\usepackage{url}

\usepackage{amssymb}

\usepackage[figuresright]{rotating}

\usepackage{xcolor}

\begin{document}

\begin{frontmatter}

\title{Latent Diffusion Modeling of Two- and Three-dimensional Fracture Networks with Sparse Observation Conditioning}

\author[UHM]{Jiawei Shen}
\author[SNL]{Hongkyu Yoon}
\author[UHM]{Jonghyun Lee\corref{cor1}}
\ead{jonghyun.harry.lee@hawaii.edu}
\cortext[cor1]{Corresponding author}

\address[UHM]{University of Hawaii at Manoa, Honolulu, HI, USA}
\address[SNL]{Sandia National Laboratories, Albuquerque, NM, USA}

\begin{abstract}

Characterizing subsurface fracture networks is challenging because fracture
geometry and connectivity are only sparsely observed and remain highly uncertain
away from measurement locations. Although stochastic discrete fracture network
(DFN) models can represent this uncertainty, conventional object-based
parameterizations are difficult to integrate efficiently with conditioning and
inverse-modeling workflows. To address this limitation, we develop a deep generative latent diffusion framework for unconditional and observation-conditioned generation of two-dimensional (2D) and
three-dimensional (3D) fracture networks. Fracture geometries and surrounding matrix are represented as binary fields and compressed into compact spatial latent representations, within which
a diffusion model learns the underlying distribution of fracture-network
structures. This latent representation provides an efficient probabilistic space for sampling geologically plausible fracture networks while enabling direct incorporation of sparse observations. In 3D cases, the fracture network is modeled directly as a
native volumetric occupancy field, while observations at discrete, multiple wells are incorporated through learned conditioning and inference-time measurement guidance.

The generated ensembles reproduce key statistical characteristics of the reference fracture networks, including fracture area/volume fraction, two-point correlation, and chord-length distributions. 
They also produce reconstructed fracture counts and gas-production responses that are comparable to those of the reference network under the considered multi-phase flow scenarios. For conditional generation, the 2D model satisfies 99.6\% of the prescribed well observations. In the more challenging 3D case, where vertical boreholes constrain full columns of the native voxel grid, the conditional realizations satisfy 84.2\% of the observed well voxels while maintaining fracture volume and geometry comparable to the reference ensemble. These results demonstrate the latent diffusion models can serve as compact, flexible probabilistic representations for fracture-network characterization, enabling both ensemble generation and sparse-observation conditioning in 2D and 3D cases.

\end{abstract}


\end{frontmatter}

\section{Introduction}

Fractures represent ubiquitous structural discontinuities within geological formations and often exert first-order control on subsurface fluid flow and the transport of mass and heat \cite{berkowitz2002characterizing,bear2012flow}. 
Accurate characterization of fracture networks is therefore important for a broad range of subsurface applications, including enhanced geothermal systems, 
unconventional hydrocarbon production \cite{karra2015effect}, geological fluid storage \cite{lackner2003guide,altman2014chemical}, groundwater and petroleum 
resource management \cite{lavoine2020discrete}, and nuclear waste disposal 
\cite{tsang2015hydrologic}. 
Numerical treatments of fractures in fractured porous media can generally be categorized into equivalent continuum models (ECMs), discrete fracture network (DFN) models, and hybrid approaches \cite{berre2019flow,ringel2021stochastic,hu2022modeling,
zhang2002numerical}. 
Continuum models represent the collective influence of fractures using effective properties defined at representative elementary volume or computational-grid scales, whereas DFN models explicitly preserve individual fracture geometries and their connectivity with surrounding matrix, which is represented as a continuum medium. Such an explicit representation facilitates the analysis of preferential flow pathways and transport behavior that are strongly influenced by fracture-network topology. In addition, interactions between fractures and the surrounding matrix can be represented more directly through fracture-matrix exchange processes, including fluid transfer, solute diffusion, heat exchange, and reactive transport across fracture surfaces \cite{yoon2017application}. As a result, DFN models are particularly useful for systems in which flow and transport are dominated by a limited number of connected fractures, where continuum upscaling may obscure critical geometric controls on permeability, breakthrough behavior, and long-term storage or recovery performance.

Despite their physical interpretability, deterministic DFN models are often 
insufficient for subsurface characterization because fracture geometries can only be partially observed and remain highly uncertain away from boreholes, cores, or exposed surfaces. 
Stochastic DFN modeling addresses this uncertainty by treating geometrical attributes such as fracture location, size, orientation, aperture, and intensity as random variables governed by prescribed probability distributions 
\cite{jing2003review,lavoine2020discrete,lei2017use}. 
For example, fracture orientations can be described by uniform, normal, or 
von Mises--Fisher distributions \cite{wood1994simulation}. 
Conventional stochastic DFN generators such as dfnWorks 
\cite{hyman2015dfnworks} construct ensembles of fracture networks by sampling 
individual fractures from specified distributions of geometry and occurrence. 
These approaches are well suited to unconditional generation, but prescribed marginal distributions of individual fracture attributes alone do not necessarily determine the higher-order spatial organization and connectivity of individual networks. For subsurface characterization, the additional challenge is therefore to condition these stochastic realizations to available observations.

Such conditioning is complicated by the different ways in which observations constrain a fracture network. Direct observations, such as fracture intersections or orientations along boreholes, provide local geometric constraints, whereas indirect hydraulic, thermal, geophysical, or production data constrain the network through its physical response and generally do not uniquely determine individual fracture geometries. Practical characterization therefore requires ensembles that both honor available observations and preserve uncertainty in poorly constrained regions.

The difficulty of conditioning depends strongly on how the fracture network is parameterized. Parsimonious and compact statistical parameterizations, such as fracture intensity and parameters controlling geometric distributions, can constrain global network properties efficiently but may not ensure that individual realizations honor local observations or measured responses \cite{chen2023fracture}. Explicit object-based parameterizations provide greater control over individual fracture geometries \cite{ringel2021stochastic}, but typically lead to variable-dimensional, discrete, and potentially multimodal parameter spaces because the number of fractures may vary among realizations.

These characteristics also complicate conventional inversion. Gradient-based methods generally require fixed-dimensional, continuous, and differentiable parameterizations, whereas object-based DFN workflows involve discrete fracture sampling, acceptance or rejection, intersection detection, connectivity changes, and repeated meshing. Small geometric changes can therefore produce abrupt changes in network connectivity and simulated flow responses, making end-to-end gradient propagation difficult. Ensemble-based data-assimilation methods avoid explicit derivatives, but covariance-based updates may become inaccurate for strongly nonlinear or multimodal parameter--observation relationships. A central challenge is therefore to develop a representation that preserves realistic fracture geometry and connectivity while providing a continuous, fixed-dimensional parameter space suitable for conditioning and uncertainty quantification.

Geostatistical simulation provides an alternative representation of spatial uncertainty. Conventional two-point approaches, such as sequential Gaussian or sequential indicator simulation, reproduce prescribed covariance or variogram structures but may fail to represent long-range connectivity and complex curvilinear 
features \cite{deutsch1998gslib}. Multiple-point statistics (MPS) methods instead extract higher-order spatial patterns from training images and can reproduce connected geological structures that cannot be adequately characterized by two-point statistics alone \cite{strebelle2002conditional,mariethoz2010direct}. 
Nevertheless, these methods generally operate directly on high-dimensional pixel- or voxel-based representations, rely on spatial kernels such as kriging \cite{yoon2008impact,yoon2012highly} and do not naturally provide a compact, continuous parameterization for inverse problems \cite{tavakoli2013comparison}. Moreover, probabilistic inversion within such high-dimensional geological representations can be computationally expensive. 
Learned generative parameterizations have therefore received increasing attention as a means of retaining complex geological structure while substantially reducing the dimension of the model space \cite{laloy2018sgan,kim2021fast}.

Deep generative models provide a promising means of representing complex fracture networks in a compact parameter space. Rather than describing a network through prescribed statistical distributions or variable-length lists of individual fractures, these models learn nonlinear representations directly from training ensembles. High-dimensional fracture fields can therefore be mapped to fixed-dimensional continuous latent variables while retaining complex geometries and connectivity. Variational autoencoders (VAEs) \cite{kingma2014autoencoding} and generative adversarial networks (GANs) \cite{goodfellow2014generative} have been applied to fracture-network characterization and conditioning \cite{chen2023fracture,deng2024characterization,ferreira2022framework}. However, strong VAE compression can smooth fine-scale structures, while GANs may suffer from unstable training and mode collapse \cite{ovanger2025statistical,dhariwal2021diffusion}.

Diffusion probabilistic models offer an alternative framework that learns the data distribution through progressive noising and learned reverse denoising \cite{ho2020denoising,dhariwal2021diffusion}. Their stable training and iterative sampling procedure make them particularly attractive for conditional generation and probabilistic inverse problems \cite{ho2022classifier,chung2023diffusion}. Latent diffusion models (LDMs) further reduce computational cost by performing diffusion in a compressed latent space \cite{rombach2022high}, and have recently been applied to geological structure generation, parameterization, data assimilation, and geophysical inversion \cite{lee2025latent,ovanger2025statistical,difederico2025latent,wang2023prior,miele2025diffusion}.

For fracture networks specifically, applications of deep generative models remain comparatively limited and have focused predominantly on 2D systems. 
Recent studies have investigated GAN-based generation of 2D fracture networks \cite{teng2025generating} and latent diffusion for 2D DFNs \cite{teng2024generative}. A three-dimensional (3D) GAN-based parameterization has also recently been proposed \cite{teng2026projection}; however, the 3D 
network is represented through separately generated 2D maps describing fracture occurrence and geometric attributes rather than through a native volumetric fracture field. Moreover, the method focuses on unconditional reproduction of fracture statistics and does not incorporate sparse field observations into the generative process. To the best of our knowledge, a latent diffusion model that directly learns a native volumetric representation of 3D fracture networks while simultaneously enabling fine-resolution conditioning on sparse borehole observations has not yet been demonstrated.

Such a representation presents additional challenges because 3D fractures are thin surfaces embedded in a volumetric domain. Their voxelized representation is therefore sparse and strongly imbalanced, and latent compression can thicken, fragment, or remove thin fracture planes. Conditional generation must consequently honor sparse observations without artificially increasing fracture density or degrading the geometry, spatial organization, and connectivity learned from the training ensemble.

To address these challenges, we develop a latent diffusion framework for unconditional and conditional generation of 2D and 3D fracture networks represented as binary occupancy fields. In 3D, fractures are modeled directly as a native volumetric field rather than reconstructed from separate 2D projections. A variational autoencoder maps the fine-resolution fracture field to a compact spatial latent representation, and an Elucidated Diffusion Model (EDM) \cite{karras2022elucidating} learns the corresponding latent distribution. For the strongly imbalanced 3D case, the reconstruction objective is modified to better preserve sparse, thin fracture structures.

Sparse observations are incorporated using complementary learned and inference-time conditioning. A conditioning encoder maps well observations and their spatial masks to the latent space, while Classifier-Free Guidance (CFG) \cite{ho2022classifier} controls their influence during sampling. Diffusion Posterior Sampling (DPS) \cite{chung2023diffusion} further enforces measurement consistency by guiding the latent trajectory using mismatch between decoded fracture fields and observations. Together, these mechanisms allow local observations to constrain generated fracture networks while retaining variability in poorly observed regions.

The main contributions of this study are threefold. First, we develop a unified latent diffusion framework for unconditional and observation-conditioned generation of fine-resolution 2D and 3D fracture networks, with 3D networks represented directly as native volumetric binary occupancy fields. Second, we introduce a conditioning strategy that combines a learned conditioning encoder, classifier-free guidance, and diffusion posterior sampling to honor sparse borehole observations while preserving variability in unconstrained regions. Third, we evaluate the generated ensembles using complementary geometric, statistical, and flow-relevant measures, including fracture volume fraction, two-point correlation functions, chord-length distributions, reconstructed fracture counts, and reservoir production responses. These contributions provide a compact probabilistic representation of complex fracture networks that reproduces key statistical and flow characteristics of the reference ensemble while enabling sparse observations to constrain local fracture occurrence.

\section{Methodology}

\subsection{Overview}
\begin{figure}[ht]
\vskip 0.2in
\begin{center}
\centerline{\includegraphics[width=\columnwidth]{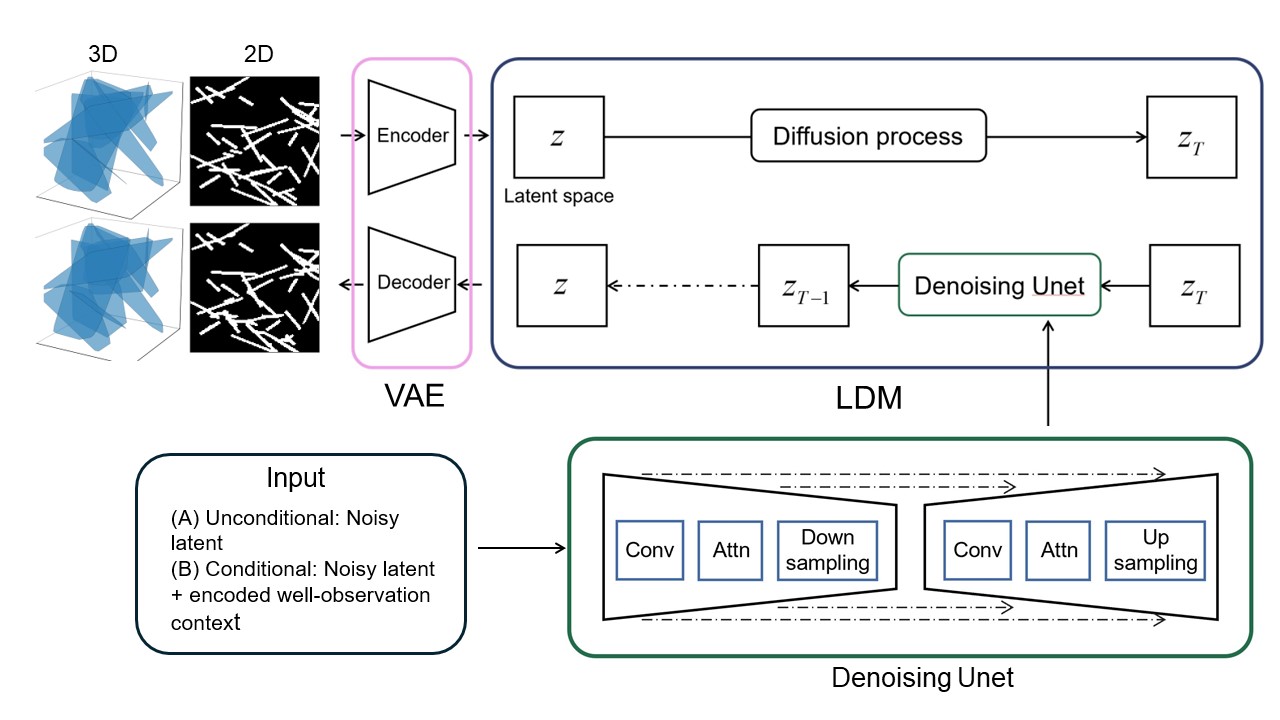}}
\caption{Schematics of the proposed latent diffusion model architecture and its inputs for fracture network generation.}

\label{Model}
\end{center}
\vskip -0.2in
\end{figure}

The proposed generator is a latent diffusion model (LDM) built from two learned components together with two inference-time conditioning mechanisms, as illustrated later in Figure \ref{Model} (the LDM box). A variational autoencoder (VAE) first compresses each fracture network into a compact, spatially structured latent space, providing a reduced-order representation that makes the subsequent diffusion process inexpensive and well-suited to downstream parameter estimation, i.e, model inversion/history matching. A diffusion model is then trained to sample from this latent space. In this work, we adopt the EDM formulation of Karras et al.~\cite{karras2022elucidating} rather than the original discrete-time Denoising Diffusion Probabilistic Model (DDPM) \cite{ho2020denoising} because its continuous-noise denoiser provides a clean estimate of the signal at every step, which we can exploit for both classifier-free guidance and hard data enforcement. Finally, the trained VAE decoder maps the generated latent space variables back to fine-resolution fracture fields with their original dimensions in 2D/3D as in Figure \ref{Model}. The same network supports both unconditional generation and conditioning on sparse well observations.

\subsection{Binary representation and network reconstruction}
\label{sec:representation}
Before compression in the latent space, each fracture network is rendered on a regular grid as a binary occupancy field, i.e., a cell is assigned to a value of one where a fracture passes through it and zero otherwise. In 2D, the fracture network is modeled as a union of thin curvilinear segments, and as shown in Figure \ref{Model} the 2D binary occupancy image represents its width of fractures in a few pixels. The autoencoder compresses in the latent space and reconstructs them onto the original image space. In 3D, each fracture is represented as a planar sheet, and we rasterize the fracture network onto the 3D grid using the same binary rule as in 2D with a voxel assigned to a value of one where a fracture surface intersects it. However, since the 3D field is far sparser than the 2D one and the sheets with only two to three voxels thick occupy a small fraction of the volume, the representation is strongly imbalanced toward the background. We address this imbalance in the loss functions (\S\ref{sec:vae}, \S\ref{sec:harddata}) rather than by changing the representation, keeping the 2D and 3D cases on a common footing.

A binary volume, whether rasterized from data or produced by the generator, stores an occupancy field rather than a list of planes, so recovering the discrete fracture network requires fitting planes back to the field. We use a local Principal Component Analysis (PCA) using gradient-free variant of sequential Random Sample Consensus (RANSAC) \cite{schnabel2007efficient}, as summarized in Algorithm~\ref{alg:reconstruct}. Because a binary field has no usable gradient, the local surface normal at each occupied voxel is estimated directly from the point cloud. Within a small neighborhood, the surface normal of a sheet structure is given by the eigenvector associated with the minimum eigenvalue of the spatial covariance matrix of neighboring occupied voxels, and the corresponding eigenvalue ratio is defined as a planarity score, for example, near one on a clean sheet and near zero inside an isotropic blob. Given these per-voxel normals, fractures are extracted one at a time by two nested loops. In the inner loop, we search over many randomly drawn seed voxels. each seed hypothesizes the plane passing through it with the seed's normal, and the hypothesis is scored by its inlier count .The occupied voxels that lie both within a distance tolerance of that plane and whose local normal is nearly parallel to the hypothesis. Then the best-scoring hypothesis is accepted as one fracture. In the outer loop, the accepted fracture's voxels are removed and the inner search repeats on the remainder, until no hypothesis reaches a minimum inlier count. Near-coplanar fractures are then merged, and each surviving fracture is finalized by refitting its normal as the smallest right-singular vector of its (mean-centered) inlier coordinates and taking the convex hull of those inliers, projected onto the fitted plane, as its polygon footprint. The output comprises a list of fracture planes, each defined by a polygon footprint, a unit normal, and a centroid. Applying identical procedures and parameters to both the generated and validation volumes ensures a consistent, unbiased basis for all plane-based comparisons.

\begin{algorithm}[htbp]
\caption{Local-PCA sequential RANSAC: reconstruct a discrete fracture network from a binary occupancy volume $B$.}
\label{alg:reconstruct}
\begin{algorithmic}[1]
\Require occupancy field $B$; for our examples, it is set to occupancy level $\ell=0.5$; PCA radius $r=2.0$; distance tolerance $\delta=2.2$; parallel tolerance $\alpha=0.90$; minimum inliers $K=150$; hypotheses per fracture $H=300$; maximum fractures $M=40$; merge tolerances $\alpha_m=0.96,\ \delta_m=3.5$ (voxel units)
\State $\mathcal{P}\gets\{\mathbf{x}:B(\mathbf{x})>\ell\}$ \Comment{occupied voxels}
\ForAll{$\mathbf{x}\in\mathcal{P}$}
  \State $\mathbf{n}_{\mathbf{x}}\gets$ smallest-eigenvalue eigenvector of $\mathrm{Cov}\big(\{\mathbf{y}\in\mathcal{P}:\lVert\mathbf{y}-\mathbf{x}\rVert\le r\}\big)$ \Comment{local sheet normal by PCA}
\EndFor
\State $\mathcal{C}\gets\varnothing$ \Comment{list of extracted fractures}
\While{$\lvert\mathcal{P}\rvert\ge K$ \textbf{and} $\lvert\mathcal{C}\rvert<M$}
  \State $\mathcal{I}^{\star}\gets\varnothing$
  \For{$H$ trials} \Comment{inner loop: hypothesis search}
    \State draw a random seed $\mathbf{s}\in\mathcal{P}$; \ $\mathbf{n}\gets\mathbf{n}_{\mathbf{s}}$, \ $c\gets\mathbf{n}\cdot\mathbf{s}$ \Comment{plane through the seed}
    \State $\mathcal{I}\gets\{\mathbf{x}\in\mathcal{P}:\lvert\mathbf{n}\cdot\mathbf{x}-c\rvert<\delta \ \textbf{and} \ \lvert\mathbf{n}_{\mathbf{x}}\cdot\mathbf{n}\rvert>\alpha\}$ \Comment{close and locally parallel}
    \If{$\lvert\mathcal{I}\rvert>\lvert\mathcal{I}^{\star}\rvert$} \ $\mathcal{I}^{\star}\gets\mathcal{I}$ \EndIf
  \EndFor
  \If{$\lvert\mathcal{I}^{\star}\rvert<K$} \ \textbf{break} \EndIf
  \State $\mathcal{C}\gets\mathcal{C}\cup\{\mathcal{I}^{\star}\}$; \ $\mathcal{P}\gets\mathcal{P}\setminus\mathcal{I}^{\star}$ \Comment{accept largest set, remove its voxels}
\EndWhile
\State merge any two fractures in $\mathcal{C}$ whose normals satisfy $\lvert\mathbf{n}\cdot\mathbf{n}'\rvert>\alpha_m$ and whose signed offsets differ by less than $\delta_m$
\ForAll{merged inlier sets $\mathcal{I}\in\mathcal{C}$}
  \State $\mathbf{c}_0\gets\mathrm{mean}(\mathcal{I})$; \ $\mathbf{n}\gets$ smallest right-singular vector of $\mathcal{I}-\mathbf{c}_0$ \Comment{SVD normal refit}
  \State \textbf{output} fracture $\big(\mathrm{convhull}(\mathcal{I}),\ \mathbf{n},\ \mathbf{c}_0\big)$
\EndFor
\end{algorithmic}
\end{algorithm}

\subsection{Latent representation via a variational autoencoder}
\label{sec:vae}
Let $\mathbf{x}$ denote a fracture field. The VAE encoder $\mathcal{E}$ maps
$\mathbf{x}$ to a compact spatial latent representation $\mathbf{z}$, and the
decoder $\mathcal{D}$ reconstructs the fine-resolution field as
$\hat{\mathbf{x}}=\mathcal{D}(\mathbf{z})$. As illustrated in Figure~\ref{Model}, a $128\times128$ 2D fracture field is encoded into a
single-channel $16\times16$ latent representation, whereas a $32 \times 32 \times 32$ 3D
fracture field is encoded into a single-channel $8 \times 8 \times 8$ latent representation,
corresponding to 256 and 512 latent variables, respectively. The spatial
structure of the latent representation preserves local geometric information
for the convolutional denoiser, while the reduced dimensionality makes the
subsequent diffusion process computationally tractable. The VAE is trained by minimizing a reconstruction term plus a weakly weighted Kullback--Leibler regularizer,
\begin{equation}
\label{eq:vae}
\mathcal{L}_{\mathrm{VAE}}
=
\mathbb{E}\!\left[
\mathcal{L}_{\mathrm{recon}}(\hat{x},x)
\right]
+
\beta_{\mathrm{KL}}
D_{\mathrm{KL}}\!\left[
q(z|x)\,\|\,\mathcal{N}(0,I)
\right].
\end{equation}
We set the regularization parameter $\beta_{\mathrm{KL}}=10^{-6}$ for the 2D model and
$\beta_{\mathrm{KL}}=10^{-4}$ for the 3D model in our examples and the chosen parameter allows the reconstruction term in Eq.~\ref{eq:vae} to be matched to the sparsity of each field. For the 2D binary images, whose fracture pixels are only moderately outnumbered by the matrix, a binary cross-entropy on the sigmoid output is adequate. For the 3D binary volumes, where the thin sheets occupy only a small fraction of the domain, a plain cross-entropy is dominated by the abundant background and tends to erase the sheets. To address the issue, we use a Tversky region-overlap loss \cite{salehi2017tversky}, which weights false negatives and false positives separately so that it can preserve thin fractures under compression. Once trained, the VAE is ``frozen'' and all subsequent diffusion training and conditioning are performed in the learned latent space, while gradients used for hard-data conditioning (\S\ref{sec:harddata}) are back-propagated through the frozen decoder without updating its parameters.

\subsection{Latent diffusion with the EDM formulation}
We model the clean latent distribution $p_{data}(\mathbf{z_0})$ with a continuous-time diffusion process, the so-called elucidated diffusion model (EDM) \cite{karras2022elucidating}. A family of noise-perturbed distributions is obtained by adding i.i.d.\ Gaussian noise of standard deviation $\sigma$,
\begin{equation}
    \mathbf{z} = \mathbf{z}_0 + \sigma \varepsilon, \varepsilon \sim \mathcal(0,\mathbf{I})
\end{equation}
then
\begin{equation}
p(\mathbf{z};\sigma)=\int p_{data}(\mathbf{z}_0)\,\mathcal{N}\!\big(\mathbf{z};\mathbf{z}_0,\sigma^2\mathbf{I}\big)\,d\mathbf{z}_0,
\end{equation}
The data is recovered as $\sigma\!\to\!0$ and for a sufficiently large $\sigma$, $\sigma_{max}$, the noise-perturbed distribution approaches an isotropic Gaussian. Samples are generated by integrating a trained approximation to the corresponding probability-flow ordinary differential equation (ODE) backward in $\sigma$, i.e, from high to low noise levels,
\begin{equation}
\label{eq:pf-ode}
\frac{d\mathbf{z}}{d\sigma}=-\sigma\nabla_{\mathbf{z}}\log p(\mathbf{z};\sigma) = \frac{\mathbf{z}-D^*(\mathbf{z};\sigma)}{\sigma} \approx \frac{\mathbf{z}-D_\theta(\mathbf{z};\sigma)}{\sigma},
\end{equation}
where $\nabla_{\mathbf{z}}\log p(\mathbf{z};\sigma)$ is the score function of the perturbed latent distribution, defined as the gradient of the log-probability density of the noised latent distribution with respect to \(z\), $D^*(\mathbf{z};\sigma)$ is the optimal denoiser that predicts the clean latent from its noised counterpart, and the denoiser \(D_\theta(z;\sigma)\) using neural networks with parameters $\theta$ approximates the optimal denoiser $D^*$. The score is expressed through the approximate denoiser \(D_\theta(z;\sigma)\). Unlike DDPM \cite{ho2020denoising}, which predicts the added noise, $D_\theta$ directly estimates $\mathbf{z}_0$; then this clean latent estimate is used by the conditioning mechanisms described below.

To keep the inputs and targets well-scaled across the wide range of $\sigma$, the denoiser is preconditioned as~\cite{karras2022elucidating}
\begin{equation}
\label{eq:edm-precond}
D_\theta(\mathbf{z};\sigma)=c_{\text{skip}}(\sigma)\,\mathbf{z}+c_{\text{out}}(\sigma)\,F_\theta\!\big(c_{\text{in}}(\sigma)\,\mathbf{z};\,c_{\text{noise}}(\sigma)\big),
\end{equation}
with $c_{\text{skip}}=\sigma_d^2/(\sigma^2+\sigma_d^2)$, $c_{\text{out}}=\sigma\sigma_d/\sqrt{\sigma^2+\sigma_d^2}$, $c_{\text{in}}=1/\sqrt{\sigma^2+\sigma_d^2}$, and $c_{\text{noise}}=\tfrac14\ln\sigma$.  Here $\sigma_d$ is the standard deviation of the latent data, and $F_\theta$ is the neural network to be trained with a U-Net architecture in this work. The model is trained using a weighted denoising objective function corresponding to denoising score matching. Noise levels are sampled from a log-normal distribution, $\ln\sigma\sim\mathcal{N}(P_{\text{mean}},P_{\text{std}}^2)$ with the mean $P_{\text{mean}}$ and standard deviation of $P_{\text{std}}$ following \cite{karras2022elucidating}. The training loss function is given by
\begin{equation}
\label{eq:edm-loss}
\mathcal{L}_{\text{EDM}}=\mathbb{E}_{\mathbf{z}_0,\sigma,\mathbf{n}}\Big[\lambda(\sigma)\,\big\|D_\theta(\mathbf{z}_0+\sigma\mathbf{n};\sigma)-\mathbf{z}_0\big\|^2\Big],\qquad \mathbf{n}\sim\mathcal{N}(\mathbf{0},\mathbf{I}),
\end{equation}
with the EDM weighting $\lambda(\sigma)=(\sigma^2+\sigma_d^2)/(\sigma\sigma_d)^2$ that equalizes the loss magnitude across noise levels. At inference, with an initial noise-perturbed latent $z^0 \sim \mathcal{N}(0, \sigma_{max}^2 \mathbf{I})$, we discretize $\sigma$ over $N$ steps from $\sigma_{\max}$ down to $\sigma_{\min}$,
\begin{equation}
\label{eq:edm-schedule}
\sigma_i=\Big(\sigma_{\max}^{1/\rho}+\tfrac{i}{N-1}\big(\sigma_{\min}^{1/\rho}-\sigma_{\max}^{1/\rho}\big)\Big)^{\rho},\qquad i=0,\dots,N-1,
\end{equation}
and integrate Eq.~\eqref{eq:pf-ode} with a second-order Euler solver with trapezoidal rule~\cite{karras2022elucidating}, which yields accurate samples in far fewer steps (we use $N=32$) than the hundreds required by typical DDPM sampling methods \cite{ho2020denoising}.

\subsection{Conditional generation}
We condition the generated fracture-network realizations on sparse well observations by combining learned conditioning with fine-resolution data-consistency mechanisms. A learned encoder maps the observed values and their location mask to a latent-resolution context for the denoiser, while classifier-free guidance controls the influence of this context during sampling. Agreement with the observations is further promoted through a training-time consistency loss in the 2D model and inference-time measurement guidance inspired by diffusion posterior sampling (DPS) in both the 2D and 3D models. These consistency mechanisms compare decoded clean-field estimates with the observations at the observed fine-grid locations, linking latent-space generation to fine-resolution constraints.
\subsubsection{Hard-data conditioning at fine resolution}
\label{sec:harddata}
In practice, well observations provide sparse constraints on the fine-grid fracture field, with each well represented by a single observed cell in 2D or a vertical column of observed cells in 3D. Both models incorporate these observations through a learned conditioning encoder, classifier-free guidance, and inference-time measurement guidance. The 2D model additionally uses a fine-resolution consistency loss during training, whereas this loss is omitted from the 3D model, as discussed below. We first describe these mechanisms and then detail their 2D and 3D implementations, including the differences in observation geometry and training objectives.

\paragraph{Learned conditioning} The sparse observations are encoded into a latent-resolution context by a small convolutional network $\mathcal{E}_\psi$. Its input places the observed values at their well locations on the fine grid (zero elsewhere) together with a mask marking the observed locations. Successive stride-2 blocks reduce $128^2\!\to\!16^2$ in 2D (and $32^3\!\to\!8^3$ in 3D), producing a context $\mathbf{c}$ that is concatenated to the noised latent at the denoiser input, giving $D_\theta(\mathbf{z};\sigma,\mathbf{c})$.

\paragraph{Classifier-free guidance} To make the conditioning strength tunable at inference, the context, along with its mask, is randomly dropped with probability $p_{\text{uncond}}$ during training, enabling a single network to learn both the conditional denoiser $D_\theta(\mathbf{z};\sigma,\mathbf{c})$ and the unconditional counterpart $D_\theta(\mathbf{z};\sigma,\varnothing)$ where $\varnothing$ represents the empty conditioning context \cite{ho2022classifier}. At sampling time, the two denoisers are linearly combined as
\begin{equation}
\label{eq:cfg}
D_\theta^{w}(\mathbf{z};\sigma,\mathbf{c})=D_\theta(\mathbf{z};\sigma,\varnothing)+w_{\text{cfg}}\big(D_\theta(\mathbf{z};\sigma,\mathbf{c})-D_\theta(\mathbf{z};\sigma,\varnothing)\big),
\end{equation}
with the classifier-free guidance (CFG) weight parameter $w_{\text{cfg}}\!\ge\!1$ amplifying the influence of the observations. $w_{\text{cfg}}=1$ recovers the plain conditional denoiser, and $w_{\text{cfg}}=0$ with a blank context recovers unconditional sampling.

\paragraph{Fine-resolution consistency loss} Because $D_\theta$ predicts the clean latent, a per-step estimate of the decoded field is directly available as $\hat{\mathbf{x}}_0=\mathcal{D}\big(D_\theta(\mathbf{z};\sigma,\mathbf{c})+\boldsymbol{\mu}\big)$, where $\boldsymbol{\mu}$ is the latent mean used for standardization. We add a consistency penalty that compares $\hat{\mathbf{x}}_0$ to the observations at the observed cells and back-propagates through the frozen decoder $\mathcal{D}$ into both $\theta$ and $\psi$, so the total training objective is given by
\begin{equation}
\label{eq:total-loss}
\mathcal{L}=\mathcal{L}_{\text{EDM}}+\lambda_{\text{cons}}\,\mathcal{L}_{\text{cons}}\big(\hat{\mathbf{x}}_0\odot\mathbf{m},\ \mathbf{y}\odot\mathbf{m}\big).
\end{equation}
Here $\mathcal{L}_{\text{cons}}$ measures the disagreement between the decoded estimate and the observations at the observed locations and $\lambda_{\text{cons}}$ is the regularization parameter. $\mathbf{m}$ is the observation location mask and $\odot$ represents Hadamard (element-wise) product. Because the fields are binary and fracture cells are sparse, we use a class-balanced binary cross-entropy; the cross-entropy is averaged separately over the observed fracture cells and the observed matrix cells and the two averages are summed, so the few fracture observations are not overwhelmed by the abundant matrix ones. The dropped (unconditional) samples have their mask blanked as well, thus the consistency term never penalizes the unconditional branch so that the unconditional sample quality is preserved.

\paragraph{Measurement guidance at inference} The learned conditioning and classifier-free guidance allow the sampling to reconstruct the observations, however, they do not guarantee exact agreement. To improve fine-resolution data consistency, we augment the sampler with a latent-space measurement correction based on diffusion posterior sampling (DPS) \cite{chung2023diffusion} at the inference stage; at every solver step, the decoded estimate is compared to the observations and the latent space is updated along the gradient that reduces the mismatch:
\begin{equation}
\label{eq:dps}
\mathbf{z}\leftarrow\mathbf{z}-w_{\text{dps}}\,\frac{\nabla_{\mathbf{z}}\big\|\big(\mathcal{D}(D_\theta(\mathbf{z};\sigma,\mathbf{c}))-\mathbf{y}\big)\odot\mathbf{m}\big\|^2}{\big\|\big(\mathcal{D}(D_\theta(\mathbf{z};\sigma,\mathbf{c}))-\mathbf{y}\big)\odot\mathbf{m}\big\|},
\end{equation}
applied at each inference step. The correction is applied once per solver step and is not multiplied by the ODE step length. The two guidance terms are complementary; classifier-free guidance steers the learned prior toward observation-consistent modes, while DPS enforces the actual measurements at each step. Their combination allows the sampler honor the observations closely while preserving realistic fracture geometry.

\paragraph{Binary well logs} We assume a well records binary occupancy of a fracture network, and the encoder input is the two channels $(\mathbf{y},\mathbf{m})$ of observed values and observed-location mask. The 2D and 3D applications differ only in the geometric footprint of a well. In 2D, a well is a single pixel; it records one value of either $1$ where the borehole intersects a fracture, or $0$ in the matrix. In 3D, a well is a vertical borehole that observes an entire depth column at a chosen $(i,j)$ footprint, so the observation is a full column of binary occupancy values (i.e., the fractures the borehole pierces and the matrix in between). The class-balanced consistency term of Eq.~\eqref{eq:total-loss} and the inference-time DPS measurement of Eq.~\eqref{eq:dps} apply unchanged. In the 2D case, learned conditioning with classifier-free guidance already places the samples close to the observations, and the inference-time DPS correction drives the remaining mismatch to essentially zero, yielding well-data satisfaction of essentially $100\%$. In the 3D case the constraint becomes stronger. A single borehole fixes a whole fine-resolution column against a compressed $8 \times 8 \times 8$ latent space, and the fracture sheets are only two to three voxels thick. It is found that in 3D cases, the training-time consistency term raises raw well satisfaction primarily by driving the global fracture density well beyond that observed in the training data and by producing unrealistically thick fracture sheets as illustrated in the next section. The featured 3D model therefore sets $\lambda_{\text{cons}}=0$ and imposes the observations through learned conditioning, classifier-free guidance, and DPS alone, whereas the 2D model, where the density cost is negligible, retains $\lambda_{\text{cons}}=0.1$. This representation-dependent trade-off is quantified in Table~\ref{tab:ablation}.

\subsection{Architecture and training settings}
The overall model architecture is illustrated in Figure \ref{Model}. The denoiser $F_\theta$ is implemented using a U-Net architecture with convolutional residual blocks, self-attention, and stride-2 down/up-sampling; the encoded well context
$\mathbf{c}$ is concatenated with the noised latent at the denoiser input. The 2D and 3D models share the same overall design, using 2D and 3D convolutions, respectively.  The encoded well context $\mathbf{c}$ is concatenated with the noised latent at the denoiser input.

We use the EDM defaults $P_{\text{mean}}=-1.2$, $P_{\text{std}}=1.2$, $\sigma_{\min}=0.002$, $\sigma_{\max}=80$, and $\rho=7$, with with $\sigma_d$ estimated from the latent statistics. Training uses the Adam optimizer with a short learning-rate warmup and gradient-norm clipping for stability, and we draw samples from an exponential moving average of the weights. For hard-data conditioning we set $p_{\text{uncond}}=0.1$; the 2D model uses a consistency weight $\lambda_{\text{cons}}=0.1$ with inference weights $w_{\text{cfg}}=3$ and $w_{\text{dps}}=4$, and the 3D model uses $\lambda_{\text{cons}}=0$ with $w_{\text{cfg}}=5$ and $w_{\text{dps}}=0.5$, both with $N=32$ solver steps during sampling. 

Detailed architectures of the VAE, EDM denoiser, and conditioning encoder, together with the complete training hyperparameters for the 2D and 3D models, are provided in Tables~S1--S4 of the Supplementary Information with a public source code repository link in the Code Availability section.

\section{Experiments and Results}

In this section, we evaluate the performance of the proposed LDM for unconditional and conditional generation of 2D and 3D fracture networks. For unconditional generation, the generated realizations are compared with held-out validation data using complementary geometric, statistical, and flow-based metrics, including fracture area/volume fraction, two-point correlation functions, chord-length distributions, reconstructed fracture counts, and reservoir production responses. We further evaluate the ability of the proposed conditional models to generate fracture-network realizations consistent with sparse well observations in both 2D and 3D.

\subsection{Fracture Network Generation}

\subsubsection{Training data description}

For the 2D case, we generate 10,000 fracture-network realizations for model training and an additional 2,000 realizations for validation. For the 3D case, 5,000 fracture-network realizations are used for model training, while another 5,000 realizations are held out from LDM training and reserved for validation.
 The 2D domain spans $128 \times 128$ m and is discretized into $128 \times 128$ pixels, whereas the 3D domain spans $10 \times 10 \times 10$ m and is discretized into $32 \times 32 \times 32$ voxels.  The geometric and stochastic parameters used to generate the 2D and 3D fracture-network datasets are summarized in Table~\ref{tab:fracture_parameters}.

The 2D fracture networks are constructed by independently sampling fracture length, orientation, and center location from the prescribed distributions listed in Table~\ref{tab:fracture_parameters}. Fracture centers are sampled uniformly at random within the interior of the domain, and individual fractures are represented as line segments on the binary grid.

The 3D fracture-network samples are generated using the dfnWorks computational suite \cite{hyman2015dfnworks}, in which the feature rejection algorithm for meshing (FRAM) \cite{hyman2014conforming} is employed to generate stochastic 3D DFNs. The domain is a cube with a side length of 10 m. The target fracture count, fracture size,
and orientation parameters are sampled according to the distributions listed in Table~\ref{tab:fracture_parameters}. The sampled $n_{\mathrm{Poly}}$ specifies the target number of fractures during DFN generation. The final number of retained fracture planes can be smaller because fractures may be rejected during DFN construction or removed during post-processing. Each generated network is subsequently rasterized onto a $32 \times 32 \times 32$ grid as a binary occupancy volume, in which a voxel is assigned a value of one if it is intersected by a fracture surface and zero otherwise. These binary occupancy fields provide the
common fine-resolution representation used by the VAE and LDM models.

 \begin{table}[htbp]
\centering
\caption{Parameters used to generate the 2D and 3D fracture-network datasets.}
\label{tab:fracture_parameters}

\begin{tabular}{llll}
\hline
\multicolumn{4}{c}{\textbf{2D fracture networks}} \\
\hline
Parameter & Minimum & Maximum & Distribution \\
\hline
Number of fractures & 30 & 60 & Discrete uniform \\
Length (pixels) & 10 & 50 & Discrete uniform \\
Orientation ($^\circ$) & 60 & 120 & Uniform \\
Aperture (pixels) & 2 & 3 & Discrete uniform \\
\hline
\end{tabular}

\vspace{0.3cm}

\begin{tabular}{lll}
\hline
\multicolumn{3}{c}{\textbf{3D fracture networks}} \\
\hline
Parameter & Value & Distribution / setting \\
\hline
Domain size & $10\times10\times10$ m & Fixed \\
Grid spacing & $0.1$ m & Fixed \\
Fracture shape & Rectangle & Fixed \\
Aspect ratio & $1.0$ & Fixed \\
Target fracture count ($n_{\mathrm{Poly}}$) & 10-19 & Discrete uniform \\
Mean orientation & $(9.52^\circ,80.82^\circ)$ & Fixed mean direction \\
Orientation concentration & $\kappa=1.0$ & Fisher distribution \\
Mean radius & $3.373$ m & Exponential distribution \\
Radius range & $1$-$50$ m & Truncated \\
\hline
\end{tabular}
\end{table}

\subsubsection{Unconditional fracture sample generation}

\begin{figure}[htbp]
\centering
\includegraphics[width=\linewidth]{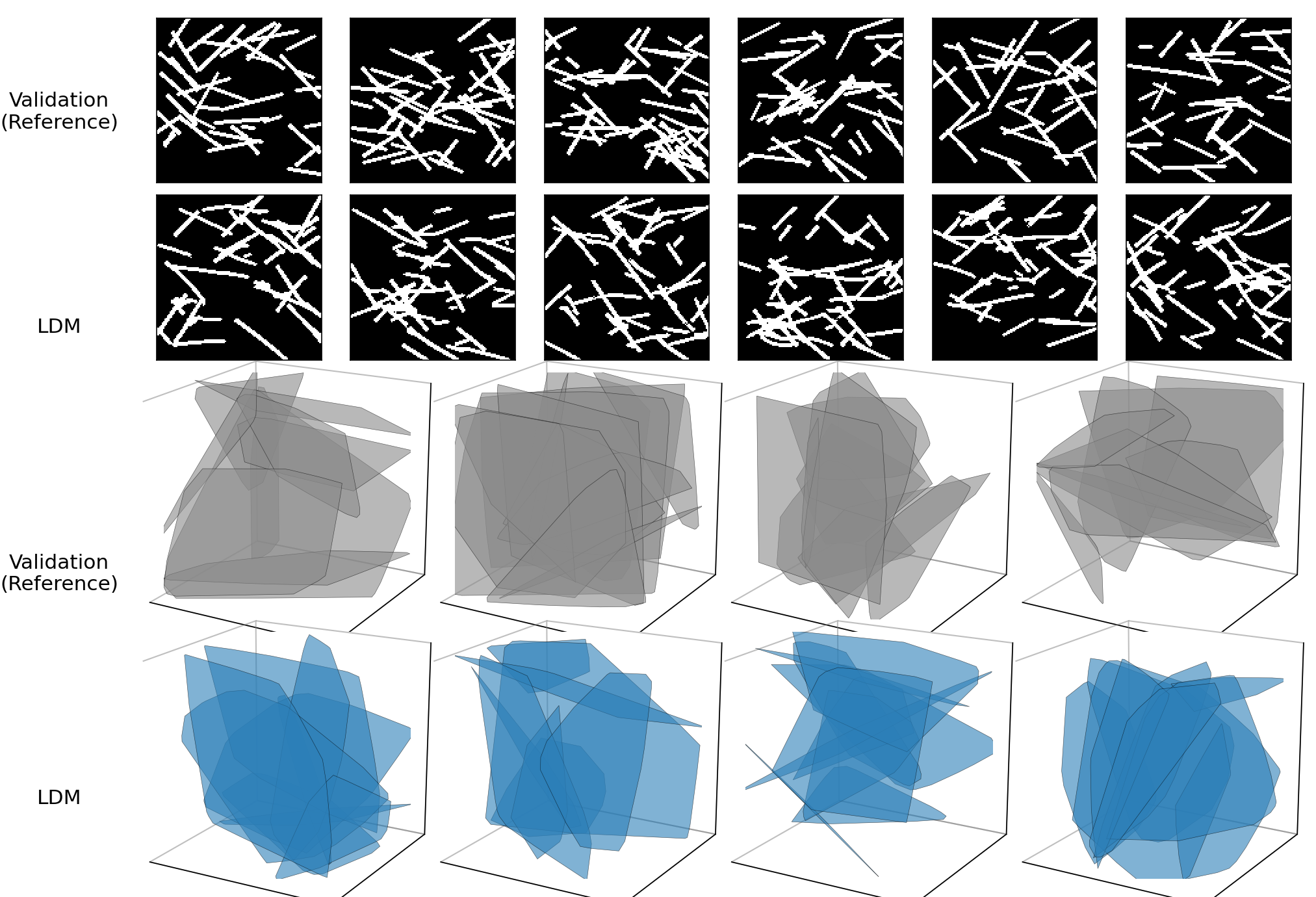}

\caption{Examples of held-out validation data and unconditional realizations sampled from the pre-trained latent diffusion models for the 2D (128 $\times 128$) and 3D ($32\times32\times32$) cases. The first and second rows show binary fracture validation and generated images in 2D, respectively, whereas the third and last two rows show reconstructed fracture planes in 3D  with validation data (grey) and generated samples (blue).}

\label{Unconditional}
\end{figure}

We train the proposed latent diffusion models to generate 2D and 3D fracture networks using latent dimensions of $16 \times 16$ and $8 \times 8 \times 8$, respectively. Note that for conditional model training, unconditional model training is not required since unconditional realizations can be generated using the same network employed for hard-data conditioning, with a blank conditioning context and $w_{\text{cfg}}=0$. The second and fourth rows of Figure~\ref{Unconditional} show randomly selected unconditional realizations generated by the LDM for the 2D and 3D cases, respectively. Qualitatively, the generated realizations reproduce the characteristic spatial structures observed in the held-out validation data, including line-like fractures in 2D and planar fracture surfaces in 3D. The following subsections provide quantitative evaluations of the generated ensembles.

\begin{figure*}[htbp]
\centering
\includegraphics[width=\linewidth]{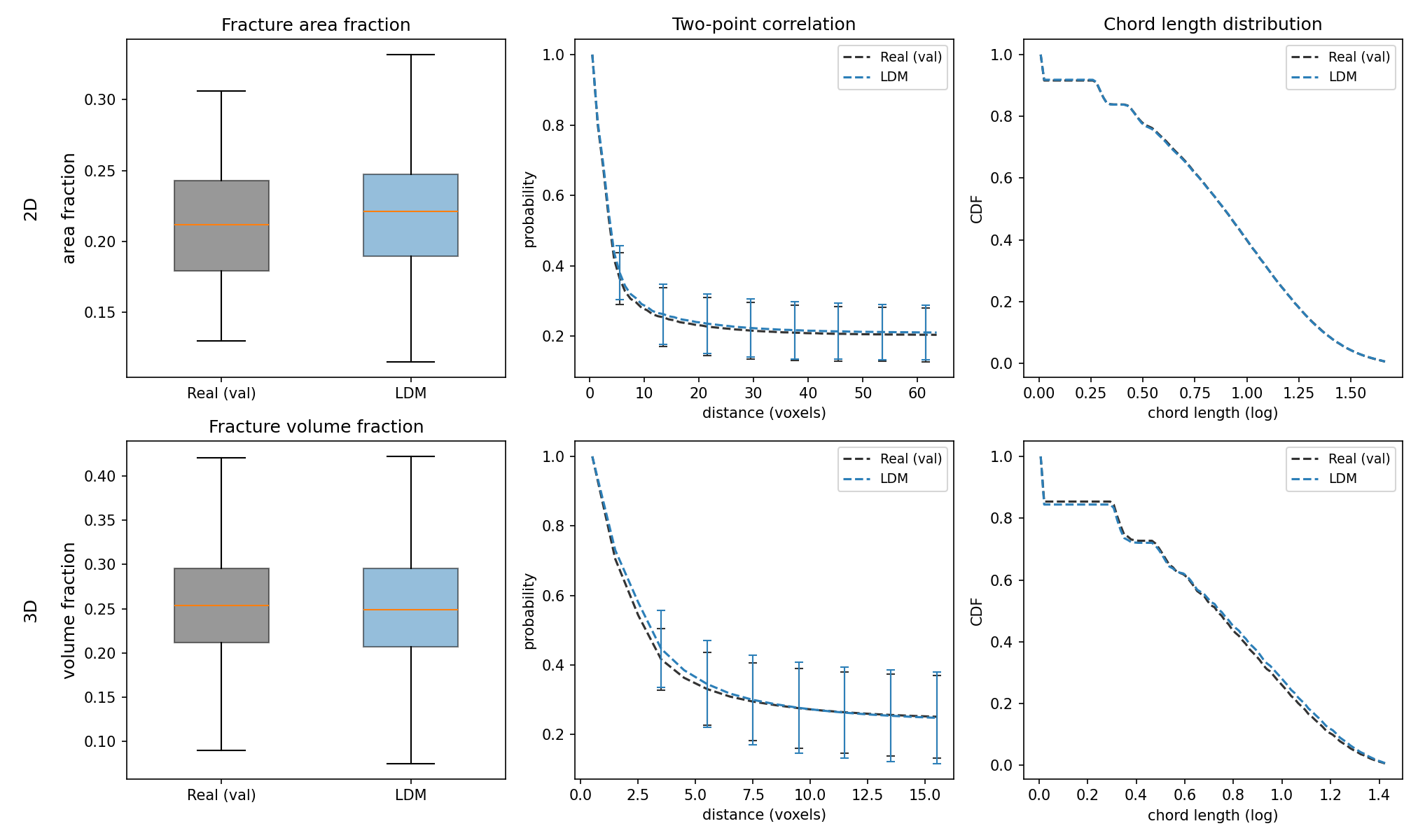}
\caption{Statistical comparison between held-out validation data (grey) and
  unconditional LDM realizations (blue) for the 2D case (top row) and the 3D   case (bottom row) using three metrics: fracture area/volume fraction (left), two-point correlation (middle; dashed lines indicate the ensemble means and   bars show the $95\%$ intervals), and chord-length cumulative distribution (right). Statistics are computed over held-out validation realizations ($2{,}000$ in 2D and $5{,}000$ in 3D) and $2{,}000$ unconditional realizations generated by the corresponding LDMs. Across all three metrics, the generated and validation distributions closely overlap, indicating that the LDM reproduces the fracture density, spatial correlation, and connectivity characteristics of the held-out validation ensemble.}
\label{Porosity 2D}
\end{figure*}

\subsubsection{Evaluation of spatial characteristics and fracture connectivity}
\label{sec:similarity}

To evaluate the spatial characteristics of the generated fracture networks, we compare the generated ensembles with held-out validation data using fracture area/volume fraction, two-point correlation functions, and chord-length distributions using PoreSpy \cite{gostick2019porespy}, a Python toolkit for microstructure analysis of porous media. The chord-length distribution provides a directional measure of the persistence and connectivity of the fracture phase. As shown in Figure~\ref{Porosity 2D}, these statistics are computed over held-out validation realizations ($2{,}000$ in 2D and $5{,}000$ in 3D) and $2{,}000$ unconditional realizations generated by the corresponding LDMs. The generated ensembles exhibit fracture area/volume fractions with mean values and variability comparable to those of the validation data. The two-point correlation functions and chord-length distributions also closely agree between the generated and validation ensembles. These results indicate that the proposed LDM reproduces the statistical and spatial characteristics of held-out validation realizations drawn from the same underlying distribution as the training data.

For the 3D case, we additionally compare the number of fracture planes recovered using the reconstruction procedure described in Algorithm~\ref{alg:reconstruct}, which provides a complementary measure of network complexity. Across the evaluated volumes, the reconstruction identifies an average of $9.98$ planes per validation volume and $10.12$ planes per generated volume, corresponding to a generated-to-reference ratio of $1.01$. This close agreement
shows that the generated 3D fracture networks successfully preserve the average number of discrete fracture planes present in the reference ensemble.

Fracture connectivity strongly influences flow and transport in fractured subsurface systems \cite{margolin1998structure, berkowitz2002characterizing}. To evaluate whether the LDM preserves flow-relevant behavior across distinct fracture-network regimes, we also perform multi-phase flow simulations with three additional 2D fracture configurations. The three cases represent (1) short fractures with high fracture density and random orientations, (2) long fractures with low fracture density and preferential orientations, and (3) long fractures with moderate fracture density and two approximately orthogonal orientation sets. Their geometric parameters are summarized in Table~\ref{tab:flow_cases}. For each configuration, a separate dataset is generated and an independent LDM is trained using the same model architecture and training procedure. From benchmark multi-phase flow simulation tests, production rates and cumulative production from the LDM-generated realizations are then compared with those from held-out validation networks drawn from the corresponding configuration.

\begin{table}[htbp]
\centering
\caption{Fracture-network configurations used for the flow-based evaluation.
Fracture lengths are sampled from discrete uniform distributions, and
orientations within prescribed angular ranges are sampled from continuous
uniform distributions.}
\label{tab:flow_cases}
\begin{tabular}{lllll}
\hline
Case & Set & Number & Length (pixels) & Orientation ($^\circ$) \\
\hline
1 & -- & 150 & $[10,20)$ & Random \\
2 & -- & 50  & $[30,50)$ & $120$--$150$ \\
3 & 1  & 50  & $[30,50)$ & $120$--$150$ \\
3 & 2  & 30  & $[30,50)$ & $30$--$60$ \\
\hline
\end{tabular}
\end{table}

\begin{figure}[htbp]
\centering
\includegraphics[width=\linewidth]{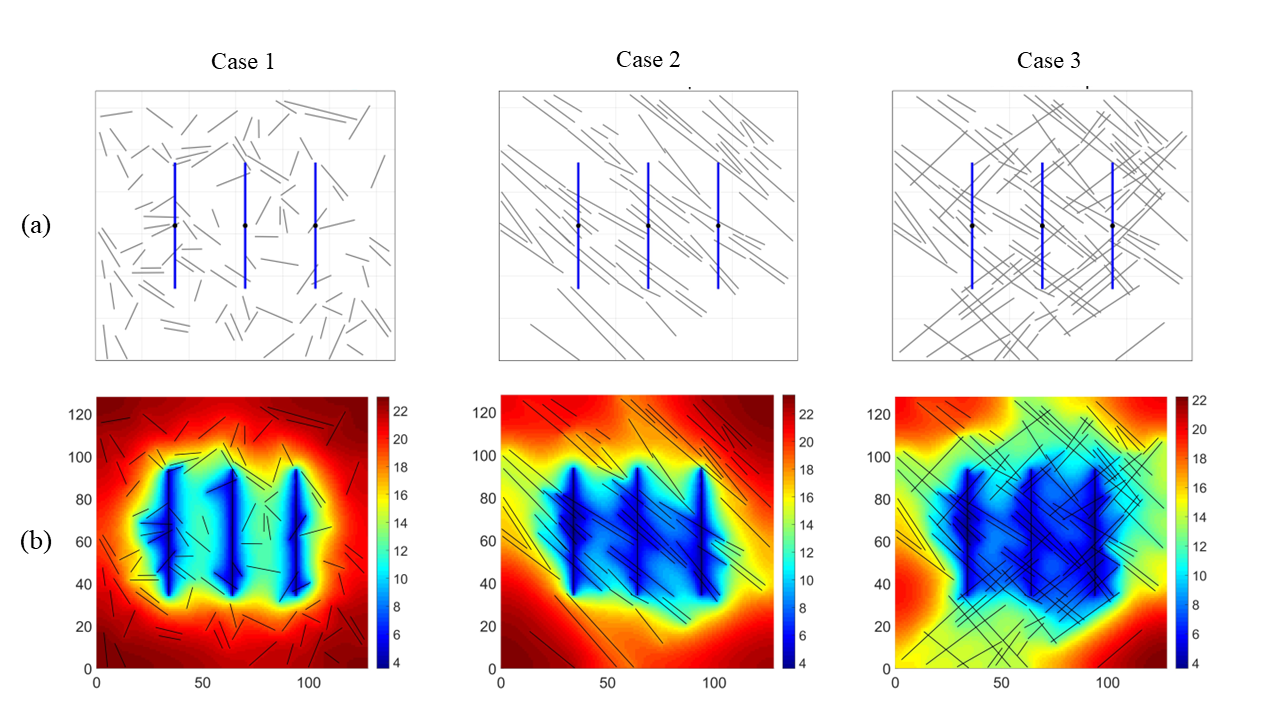}

\caption{(a) Representative fracture configurations for the three simulation cases. The hydraulic fractures (blue) are identical across the three cases, whereas the natural fracture networks (gray) vary. (b) Corresponding pressure distributions after 10 years of production.}
\label{OpenShale_scheme}
\end{figure}

Representative fracture configurations for the three cases are shown in Figure~\ref{OpenShale_scheme} (a). All other reservoir and well configurations are kept fixed so that differences in the production responses arise primarily from variations in the natural fracture networks. In particular, three parallel hydraulic fractures are placed near the center of the domain with a spacing of 30 m.

The reservoir-simulation parameters are summarized in Table~\ref{Openshale Setup}. Simulations are performed using the MATLAB Reservoir Simulation Toolbox (MRST) OpenShale Module \cite{wang2020mrst}, which requires fractures to be represented in discrete fracture network (DFN) format. The Line Segment Detector (LSD) algorithm \cite{von2012lsd} is therefore used to extract individual fractures from the binary LDM realizations.

\begin{table}[htbp]
\centering
\caption{Reservoir-simulation parameters.}
\label{Openshale Setup}
\begin{tabular}{ccc}
\toprule
Property & Unit & Value \\
\midrule
Domain dimensions & $\mathrm{m}^2$ & $128 \times 128$ \\
Formation thickness & m & 90 \\
Initial reservoir pressure & MPa & 24 \\
Natural fracture permeability & mD & 0.01 \\
Hydraulic fracture permeability & mD & 0.1 \\
Fracture aperture & m & 0.003 \\
Matrix porosity & -- & 0.03 \\
Matrix permeability & nD & 20 \\
Well BHP & MPa & 3.45 \\
\bottomrule
\end{tabular}
\end{table}

\begin{figure}[htbp]
    \centering
   \includegraphics[width=\linewidth]{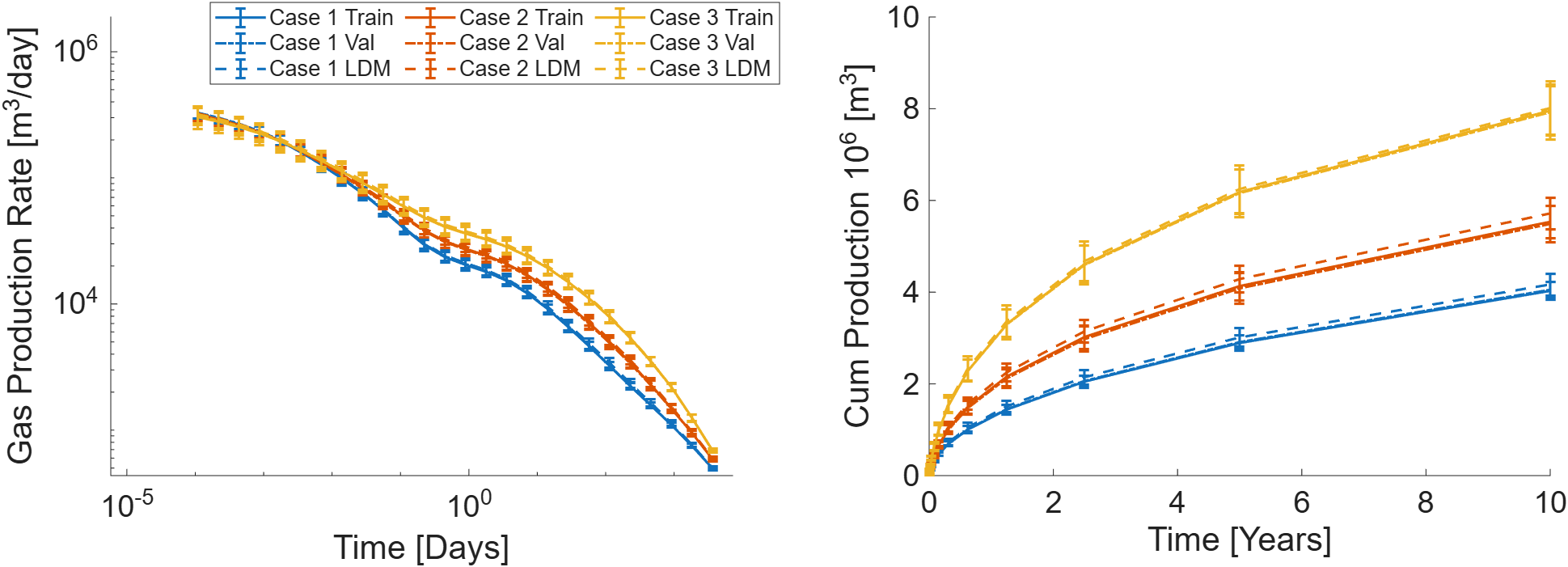}
    \caption{Comparison of gas-production responses for the training and
    validation fracture-network ensembles and the realizations generated by
    the independently trained, case-specific LDMs for the three fracture
    configurations. The left panel shows gas production rate, and the right
    panel shows cumulative gas production. Error bars indicate $\pm 1$ standard deviation across 100 realizations in each ensemble.}
    \label{OpenShale}
\end{figure}

Figure~\ref{OpenShale} compares the gas production rates and cumulative
production obtained from the training and validation fracture-network
ensembles with those from the corresponding LDM-generated realizations for the three considered cases. For each case, the gas-production analysis was conducted using 100 realizations from each of the training, validation, and LDM-generated ensembles.. Across all three fracture configurations, the training and validation ensembles exhibit closely aligned production-rate trends and cumulative-production behavior, indicating that the validation data are representative of the corresponding training distributions in terms of flow response. The LDM-generated ensembles closely reproduce these responses, with comparable overall magnitudes and temporal evolution and substantial overlap in the corresponding variability ranges. Moreover, the distinct separation in production behavior among the three fracture
configurations is consistently preserved in the generated realizations. These results indicate that the generated fracture networks reproduce important flow-relevant characteristics of the corresponding validation ensembles.

\subsection{Conditional Fracture Network Generation}

\subsubsection{2D realizations conditioned on hard data}
\label{sec:harddata_results}

Sparse observations collected at wells can be used to constrain the fracture-network distribution and reduce uncertainty near the observation locations. The well data considered here consist of sparse point observations defined on the fine-resolution grid. We therefore impose these observations directly at the $128\times128$ resolution using the learned conditioning encoder, classifier-free guidance, and the fine-resolution consistency loss described in \S\ref{sec:harddata}, with diffusion posterior sampling (DPS) further enforcing measurement consistency during inference. We demonstrate the approach using $N_w=9$ wells sampled from a held-out validation realization, thereby evaluating generalization to a fracture network that was not used during model training. Each well provides a binary observation, with a value of $1$ when the borehole intersects a fracture and $0$ otherwise.

\begin{figure}[htbp]
\vskip 0.2in
\begin{center}
\centerline{\includegraphics[width=\columnwidth]{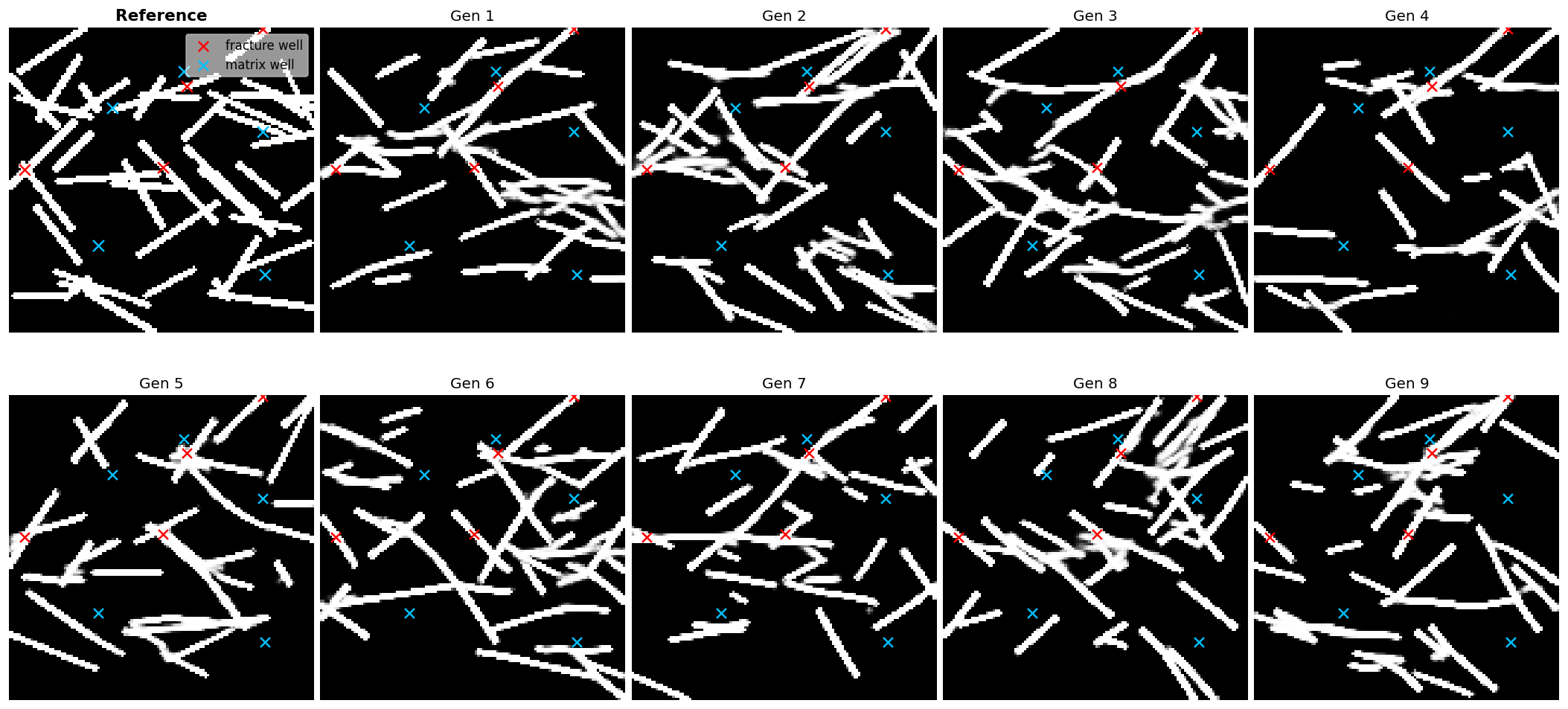}}
\caption{Fine-resolution ($128\times128$) conditional realizations constrained by discrete hard data at nine well locations. The held-out validation reference network (top left) is followed by nine generated realizations. Red crosses denote fracture wells and cyan crosses denote matrix wells. The generated realizations closely honor the well observations while retaining variability away from the observation locations.}
\label{fig:hd_realizations}
\end{center}
\vskip -0.2in
\end{figure}

Figure~\ref{fig:hd_realizations} shows the held-out reference network together with nine conditional realizations. Fracture wells (red crosses) and matrix wells (cyan crosses) are consistently honored across the generated realizations; in most cases, generated fractures pass through the fracture-well locations, whereas matrix-well locations remain fracture-free. At the same time, the fracture geometry away from the observation locations varies substantially among realizations. This behavior is consistent with the intended role of sparse conditioning. The observations constrain fracture occurrence locally while allowing the generator to represent the remaining uncertainty in poorly constrained regions based on the prior fracture network distribution.

To quantify observation consistency, each realization is binarized using a threshold of $0.5$ and compared with the well observations over $100$ conditional realizations. Fracture-well and matrix-well satisfaction rates are reported separately to prevent the less frequent fracture observations from being obscured by the more abundant matrix observations. As summarized in Table~\ref{tab:hd_satisfaction}, the conditional realizations satisfy the hard data with high accuracy; fracture-well satisfaction is $99.2\%$ ($397$ out of $400$ pixels), matrix-well satisfaction is $99.8\%$ ($499$ out of $500$ pixels), and the overall agreement is $99.6\%$ ($896$ out of $900$ pixels). Learned conditioning together with classifier-free guidance ($w_{\text{cfg}}=3$) already enforces the samples toward observation-consistent realizations, while the inference-time DPS correction ($w_{\text{dps}}=4$, $N=32$ steps) further reduces the remaining mismatch.

\begin{table}[htbp]
\centering
\caption{2D hard-data satisfaction over $100$ conditional realizations (binarization threshold $0.5$) for $N_w=9$ wells ($4$ fracture wells and $5$ matrix wells). Satisfaction denotes the fraction of well pixels for which the generated fracture/matrix label agrees with the corresponding observation.}
\label{tab:hd_satisfaction}
\begin{tabular}{lcc}
\hline
Well type & Matches / total & Satisfaction \\
\hline
Fracture wells & $397/400$ & $99.2\%$ \\
Matrix wells & $499/500$ & $99.8\%$ \\
Overall & $896/900$ & $99.6\%$ \\
\hline
\end{tabular}
\end{table}

\subsubsection{3D realizations conditioned on hard data}
\label{sec:harddata3d}

The same conditioning framework extends to the native volumetric 3D case, where each well is represented as a vertical borehole that records a full depth column of binary occupancy at a given horizontal grid location $(i,j)$. This imposes a substantially stronger constraint than a single 2D well pixel. Each borehole constrains an entire fine-resolution column through a compressed $8 \times 8 \times 8$ latent representation, while the fracture surfaces to be reconstructed are only two to three voxels thick. We condition on $N_w=2$ boreholes sampled from a held-out evaluation volume and generate realizations using $w_{\rm cfg}=5$, $w_{\rm dps}=0.5$, $\lambda_{\rm cons}=0$, and $N=32$ steps.

\begin{figure}[h!]
\vskip 0.2in
\begin{center}
\centerline{\includegraphics[width=\columnwidth]{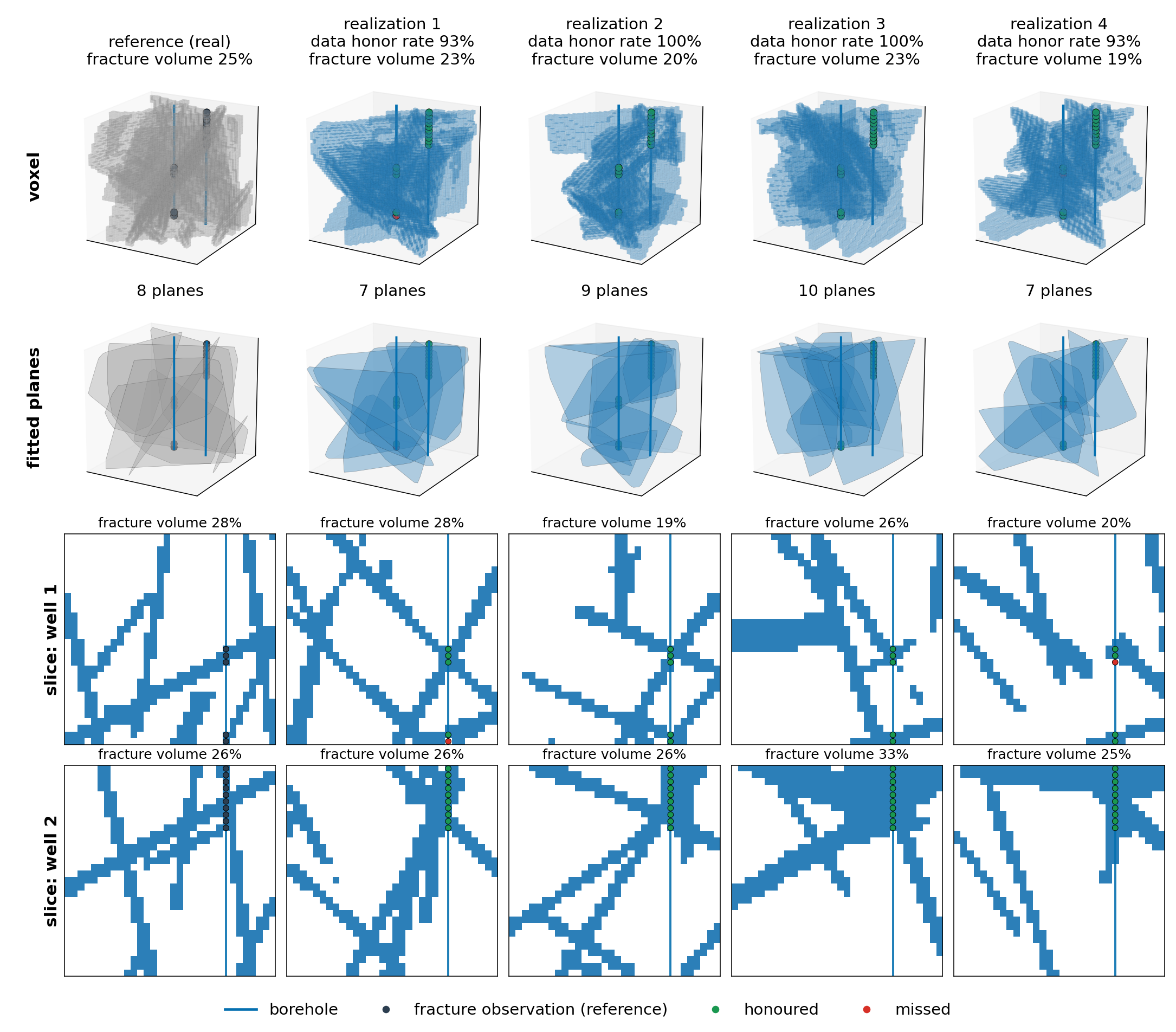}}
\caption{Native volumetric ($32 \times 32 \times 32$) conditional realizations constrained by two vertical-borehole logs. Columns show a representative held-out validation reference volume, followed by four conditional realizations selected from a larger ensemble generated for that reference. Rows show the raw binary occupancy field (voxel), reconstructed discrete fractures (fitted planes), and raw-occupancy cross-sections along the two boreholes (slice). Boreholes are indicated by solid blue lines, and fracture observations along the boreholes are marked according to whether they are satisfied (green) or missed (red) by each realization.}
\label{fig:hd3d}
\end{center}
\vskip -0.2in
\end{figure}

Because the raw binary occupancy of a 3D fracture network can be difficult to interpret visually, particularly at fracture volume fractions of approximately $25\%$, we present each realization using three complementary representations as shown in Figure~\ref{fig:hd3d}; (1) the raw binary occupancy field (voxel), (2) the discrete fractures reconstructed using Algorithm~\ref{alg:reconstruct} (fitted planes), and (3) raw-occupancy cross-sections along the two boreholes (slice). The reconstructed-plane view provides an interpretable representation of the inferred fracture geometry, whereas the cross-sectional views retain the original voxel-scale occupancy and directly show whether the borehole observations are satisfied. These representations are therefore intended to be interpreted together. The held-out validation volume is shown alongside four conditional realizations selected from a larger generated ensemble.

To quantify observation consistency across multiple reference networks, we evaluate $10$ held-out validation volumes with $10$ conditional realizations generated for each volume with its well observations, yielding $100$ realizations in total. Well observations are pooled in the same manner as in the 2D case. Because each vertical borehole records a full depth column, satisfaction is evaluated at the level of individual well voxels; a fracture voxel is considered satisfied when the generated occupancy exceeds the binarization threshold, whereas a matrix voxel is satisfied when it remains below the threshold. As summarized in Table~\ref{tab:hd3d_satisfaction}, the conditional realizations satisfy $86.2\%$ of the fracture-well voxel reconstruction ($2120$ out of $2460$ voxels) and $82.9\%$ of the matrix-well voxel reconstruction ($3267$ out of $3940$ voxels), corresponding to an overall satisfaction rate of $84.2\%$ ($5387$ out of $6400$ voxels).

To quantify the effect of observation conditioning, we additionally evaluate
unconditional realizations generated by the same model with the conditioning
context removed against the same borehole observations. The corresponding
unconditional satisfaction ratios are $26.7\%$ ($657$ out of $2460$ voxels)
for fracture-well voxels, $71.2\%$ ($2805$ out of $3940$ voxels) for
matrix-well voxels, and $54.1\%$ ($3462$ out of $6400$ voxels) overall.
These values are substantially lower than the conditional results, particularly
for the fracture-well voxels. Meanwhile, the fracture volume fraction increases
slightly from $25.9\%$ for the unconditional realizations to $26.8\%$
for the conditional realizations, with both values remaining close to the
validation-ensemble value of $25.4\%$.

Figure~\ref{fig:hd3d} provides a qualitative illustration using a representative held-out validation volume whose well-satisfaction rate is close to the ensemble average, together with visualization of four realizations from its generated ensemble. All quantitative statistics reported above including the well voxel reconstruction or data honor rate and fracture volume ratio are computed over the complete evaluation set. The lower satisfaction rate in 3D relative to the 2D case is possibly due to the substantially stronger conditioning constraint on the well voxel types; each 3D borehole specifies a complete fine-resolution column with spare fracture surface intersects in a compressed $8 \times 8 \times 8$ latent space, while the fracture surfaces themselves remain only a few voxels thick. Importantly, the mean fracture volume fraction of the conditional realizations, $26.8 \%$ (with a std of $5.7\%$), remains close to the training-ensemble mean of $25.3\%$ (with a std of $5.9\%$), indicating that the improved observation consistency is achieved without substantially altering the overall fracture density.

\begin{table}[htbp]
\centering
\caption{3D hard-data satisfaction evaluated over $10$ held-out validation reference volumes with $10$ conditional realizations generated for each reference ($100$ realizations in total; binarization threshold $0.5$; $w_{\text{cfg}}=5$, $w_{\text{dps}}=0.5$, $\lambda_{\text{cons}}=0$, and $N=32$ steps). Each vertical borehole records a full depth column, and satisfaction is therefore pooled over individual well voxels. A fracture voxel is considered matched when the generated occupancy exceeds the threshold, whereas a matrix voxel is matched when it remains below the threshold.}
\label{tab:hd3d_satisfaction}
\begin{tabular}{lcc}
\hline
Well log type & Matches / Total voxels & Satisfaction ratio\\
\hline
Fracture voxel & $2120/2460$ & $86.2\%$ \\
Matrix voxl& $3267/3940$ & $82.9\%$ \\
Overall & $5387/6400$ & $84.2\%$ \\
\hline
\end{tabular}
\end{table}

\paragraph{Consistency-loss ablation}
The observation-consistency loss $\mathcal{L}_{\rm cons}$ in Eq.~\eqref{eq:total-loss} is applied during training to encourage the decoded fracture field to match the prescribed observations at the observed locations.
Its effect differs markedly between the 2D and 3D cases. As shown in Table~\ref{tab:ablation}, setting $\lambda_{\text{cons}}=0.1$ in 2D increases the overall well-satisfaction rate from $94.3\%$ to $99.4\%$, while the generated fracture area fraction changes only modestly from $21.0\%$ to $19.5\%$, compared with a training-ensemble mean of $22.9\%$. We therefore retain $\lambda_{\text{cons}}=0.1$ for the featured 2D model.

In 3D, however, increasing $\lambda_{\text{cons}}$ from $0$ to $0.1$
improves the overall well-satisfaction rate from $84.2\%$ to $91.7\%$, but
also increases the mean generated fracture volume fraction from $26.8\%$ to
$46.6\%$, substantially above the training-ensemble mean of $25.3\%$. Thus,
the improvement in observation consistency is accompanied by a pronounced
over-representation of the fracture phase. For this reason, we use
$\lambda_{\text{cons}}=0$ for the featured 3D model.

\begin{table}[htbp]
\centering
\caption{Ablation of the training-time consistency weight
$\lambda_{\text{cons}}$. Inference parameters are fixed at the operating
point used for each case. For the 2D case ($128^2$), $10$ reference networks
with $20$ realizations per reference are evaluated using
$w_{\text{cfg}}=3$ and $w_{\text{dps}}=4$. For the 3D case ($32^3$),
$10$ reference volumes with $10$ realizations per reference are evaluated
using $w_{\text{cfg}}=5$ and $w_{\text{dps}}=0.5$. Boldface indicates the
setting used in the featured model.}
\label{tab:ablation}
\begin{tabular}{llcc}
\hline
Case & $\lambda_{\text{cons}}$ & Overall satisfaction & Gen.\ fracture fraction \\
\hline
\multirow{2}{*}{2D ($128^2$)}
& $0$ & $94.3\%$ & $21.0\%$ \\
& $\mathbf{0.1}$ & $\mathbf{99.4\%}$ & $\mathbf{19.5\%}$ \\
\hline
\multirow{2}{*}{3D ($32^3$)}
& $\mathbf{0}$ & $\mathbf{84.2\%}$ & $\mathbf{26.8\%}$ \\
& $0.1$ & $91.7\%$ & $46.6\%$ \\
\hline
\end{tabular}
\end{table}

\subsection{Computational cost}

Table~\ref{tab:timing} reports the wall-clock cost of generating batches of realizations using the $N=32$-step EDM sampler. For the 2D model, unconditional sampling requires approximately $11$--$17$~ms per $128\times128$ realization. Incorporating classifier-free guidance and inference-time DPS increases the cost to approximately $50$--$65$~ms per conditional realization, primarily because conditioning requires two denoiser evaluations together with gradient back-propagation through the decoder at each sampling step.

The 3D model incurs a higher per-realization cost because of the increased dimensionality of the volumetric representation. Unconditional sampling of a $32 \times 32 \times 32$ grid realization requires approximately $190$--$210$~ms, whereas conditional sampling requires approximately $0.7$--$0.8$~s per realization because DPS additionally requires repeated back-propagation through the 3D decoder. Nevertheless, an ensemble of conditioned 3D realizations can be generated within tens of seconds at the batch sizes considered here. In both 2D and 3D, the per-realization cost decreases with increasing batch size because fixed computational overheads are amortized across the batch. All timings were measured on a single NVIDIA RTX~A6000 GPU with Dual Intel Xeon 2.1 GHz 32 core CPUs and 128 GB RAM.

\begin{table}[htbp]
\centering
\caption{Wall-clock sampling times measured on a single NVIDIA RTX~A6000 GPU using $N=32$ solver steps. ``Cond.'' denotes conditional sampling using classifier-free guidance and DPS ($w_{\text{cfg}}=3$, $w_{\text{dps}}=4$ in 2D; $w_{\text{cfg}}=5$, $w_{\text{dps}}=0.5$ in 3D), whereas ``Uncond.'' denotes unconditional prior sampling. Per-realization time is obtained by dividing the total batch time by the batch size. All reported timings use deterministic cuDNN settings and correspond to the minimum of three repetitions.}
\label{tab:timing}
\begin{tabular}{llccc}
\hline
Case & Mode & Batch & Total time & Per realization \\
\hline
2D ($128^2$) & Uncond. & $100$ & $1.7$~s & $17$~ms \\
2D ($128^2$) & Uncond. & $200$ & $2.2$~s & $11$~ms \\
2D ($128^2$) & Cond.\ (CFG+DPS) & $100$ & $6.5$~s & $65$~ms \\
2D ($128^2$) & Cond.\ (CFG+DPS) & $200$ & $9.9$~s & $50$~ms \\
3D ($32^3$) & Uncond. & $32$ & $6.7$~s & $210$~ms \\
3D ($32^3$) & Uncond. & $64$ & $12.0$~s & $188$~ms \\
3D ($32^3$) & Cond.\ (CFG+DPS) & $32$ & $24.5$~s & $764$~ms \\
3D ($32^3$) & Cond.\ (CFG+DPS) & $64$ & $45.2$~s & $706$~ms \\
\hline
\end{tabular}
\end{table}

\section{Conclusion}

This study demonstrates that the latent diffusion model provides a flexible deep generative framework for learning complex fracture-network distributions in
both 2D and 3D. 
Fracture networks constitute a particularly challenging generative setting because they are composed of sparse, thin geometric structures whose connectivity and spatial organization must be preserved during compression and generation. This challenge is further amplified in 3D, where fractures occupy only a small fraction of the volumetric domain.
By combining a common binary occupancy representation with VAE-based latent compression and diffusion-based generative modeling, the proposed framework is able to learn and generate both line-like 2D fracture structures and native volumetric 3D fracture surfaces within a compact latent space.

The generated ensembles were evaluated using a range of geometric, statistical, and flow-relevant measures, including fracture area/volume fraction, two-point correlation functions, chord-length distributions, reconstructed fracture counts, and reservoir production responses. Across these measures, the LDM-generated realizations reproduce the principal characteristics of the corresponding validation ensembles. The reservoir simulations further show that the generated and validation ensembles exhibit comparable gas-production-rate trends and cumulative-production behavior across the considered fracture configurations, indicating that the generated fracture networks preserve important flow-relevant characteristics.

Beyond unconditional generation, the framework demonstrates how deep generative models can be coupled with sparse physical observations through learned conditioning, classifier-free guidance, and diffusion posterior sampling. 
In the 2D case, the conditional realizations satisfy 99.6\% of the prescribed well observations while retaining variability away from the observation locations. Conditioning is substantially more demanding in 3D, where each vertical borehole constrains a full fine-resolution column within a compressed latent representation and the target fracture surfaces remain thin and sparse. 
Nevertheless, the 3D conditional realizations satisfy 84.2\% of the observed well voxels while maintaining an overall fracture volume fraction close to that of the reference ensembles. 
Overall, these results demonstrate the potential of latent diffusion models as compact probabilistic representations for generating fracture-network ensembles and incorporating sparse local observations in subsurface characterization and uncertainty quantification applications.

\section{Acknowledgements}
This opportunity is supported by the National Science Foundation (NSF) Strengthening the CyberInfrastructure Professionals Ecosystem (SCIPE) Cyberinfrastructure Pacific Professionals (CI-PP) grant number 2417946. HY and JL were also partly supported by the US DOE project - the Science-informed Machine Learning for Accelerating Real-Time Decisions in Subsurface Applications (SMART) Initiative and HY was partly supported by the U.S. DOE Science Foundations for Energy Earthshots (SFEE) project, Advanced Multi-Physics Machine Learning for Subsurface Energy Systems Across Scales under Award No. DE-SC0024703. JS was partly supported by the U.S. DOE SFEE project, Ka mana o ka lā: Modeling our energy future under Award No. DE-SC0024728. Sandia National Laboratories is a multimission laboratory managed and operated by National Technology \& Engineering Solutions of Sandia, LLC, a wholly owned subsidiary of Honeywell International Inc., for the U.S. Depart-ment of Energy’s National Nuclear Security Administration under contract DE-NA0003525. This written work is authored by an employee of NTESS. The employee, not NTESS, owns the right, title and interest in and to the written work and is responsible for its contents. Any subjective views or opinions that might be expressed in the written work do not necessarily represent the views of the U.S. Government. The publisher acknowledges that the U.S. Government retains a non-exclusive, paid-up, irrevocable, world-wide license to pub-lish or reproduce the published form of this written work or allow others to do so, for U.S. Government purposes. The DOE will provide public access to results of federally sponsored research in accordance with the DOE Public Access Plan.

\section*{Code availability}

The source code used for model training, conditional generation, and
reproduction of the main numerical results is publicly available at
\url{https://github.com/jiaweish23/fracture-latent-diffusion}.

\section*{Supplementary Information}

% Supplementary table and figure numbering
\setcounter{table}{0}
\renewcommand{\thetable}{S\arabic{table}}

\setcounter{figure}{0}
\renewcommand{\thefigure}{S\arabic{figure}}

% ============================================================
\subsection*{S1. Network architectures}
% ============================================================

The framework uses three learned networks in each spatial dimension:
a variational autoencoder (VAE) that compresses the binary fracture field
into a compact spatial latent representation (Table~\ref{tab:supp_vae}),
an EDM denoiser operating in the latent space
(Table~\ref{tab:supp_denoiser}), and a conditioning encoder that maps
sparse well observations to a latent-resolution context
(Table~\ref{tab:supp_condencoder}).
The 2D and 3D models share the same overall architectural design while
using dimension-specific convolutional operators, channel widths, and
numbers of resolution levels. Convolutional blocks use GroupNorm and
SiLU activations. The denoiser additionally employs weight-standardized
convolutions and a sinusoidal noise-level embedding followed by a
two-layer multilayer perceptron (MLP).

% ------------------------------------------------------------
% Table S1: VAE
% ------------------------------------------------------------

\begin{table}[htbp]
\centering
\small
\caption{
Architecture of the variational autoencoder (VAE).
The binary occupancy field is compressed into a single-channel spatial
latent representation. The 2D and 3D encoders first embed the binary
input through a learned two-entry embedding of width \texttt{dim}, followed
by convolutional blocks with stride-2 downsampling. The decoder mirrors
the encoder to reconstruct the fine-resolution binary field.
}
\label{tab:supp_vae}

\begin{tabular}{lll}
\hline
Component & 2D VAE & 3D VAE \\
\hline
Input field
& $128\times128\times1$
& $32\times32\times32\times1$ \\

Base width (\texttt{dim})
& $32$
& $96$ \\

Channel multipliers
& $(1,\,2,\,4)$
& $(1,\,2)$ \\

Stride-2 downsampling steps
& $3$ ($128\rightarrow16$)
& $2$ ($32\rightarrow8$) \\

Latent representation
& $1\times16\times16$ ($256$ variables)
& $1\times8\times8\times8$ ($512$ variables) \\

Output activation
& Sigmoid
& Sigmoid \\

Number of parameters
& $2.35$ M
& $16.42$ M \\
\hline
\end{tabular}
\end{table}

% ------------------------------------------------------------
% Table S2: EDM denoiser
% ------------------------------------------------------------

\begin{table}[htbp]
\centering
\small
\caption{
Architecture of the EDM denoiser.
The denoiser is a U-Net composed of residual convolutional blocks,
linear attention at each resolution, and full self-attention at the
bottleneck. For conditional generation, the single-channel context
volume is concatenated with the noised latent at the input. A zero context
is used for the unconditional branch during classifier-free-guidance
training and for unconditional sampling.
}
\label{tab:supp_denoiser}

\begin{tabular}{lll}
\hline
Component & 2D denoiser & 3D denoiser \\
\hline
Input (latent $\oplus$ context)
& $2\times16\times16$
& $2\times8\times8\times8$ \\

Base width (\texttt{dim})
& $128$
& $128$ \\

Channel multipliers
& $(1,\,2,\,4)$
& $(1,\,2,\,4,\,8)$ \\

Residual blocks per resolution
& $2$
& $2$ \\

Attention at each resolution
& Linear attention
& Linear attention \\

Bottleneck attention
& Full self-attention
& Full self-attention \\

Noise-level embedding
& Sinusoidal + 2-layer MLP
& Sinusoidal + 2-layer MLP \\

Skip connections
& U-Net concatenation
& U-Net concatenation \\

Output
& $1\times16\times16$
& $1\times8\times8\times8$ \\

Number of parameters
& $35.20$ M
& $385.96$ M \\
\hline
\end{tabular}
\end{table}

% ------------------------------------------------------------
% Table S3: Conditioning encoder
% ------------------------------------------------------------

\begin{table}[htbp]
\centering
\small
\caption{
Architecture of the conditioning encoder $\mathcal{E}_{\psi}$.
The encoder receives two fine-resolution channels containing the observed
values and the corresponding observation mask. Stride-2 convolutional
blocks reduce the spatial resolution to that of the latent representation,
followed by a pointwise projection to a single-channel context
($1\times1$ in 2D and $1\times1\times1$ in 3D).
The 3D encoder additionally contains a stride-1 refinement block.
}
\label{tab:supp_condencoder}

\begin{tabular}{lll}
\hline
Component & 2D encoder & 3D encoder \\
\hline
Input
& $2\times128\times128$
& $2\times32\times32\times32$ \\

Stride-2 blocks
& $3$ ($128\rightarrow64\rightarrow32\rightarrow16$)
& $2$ ($32\rightarrow16\rightarrow8$) \\

Channel schedule
& $2\rightarrow32\rightarrow64\rightarrow32\rightarrow1$
& $2\rightarrow32\rightarrow64\rightarrow32\rightarrow1$ \\

Normalization / activation
& GroupNorm(8), SiLU
& GroupNorm(8), SiLU \\

Output context
& $1\times16\times16$
& $1\times8\times8\times8$ \\

Number of parameters
& $37.9$ k
& $112.7$ k \\
\hline
\end{tabular}
\end{table}

% ============================================================
\subsection*{S2. Training and sampling settings}
% ============================================================

The VAE is trained first and subsequently frozen. The latent diffusion
model and conditioning encoder are then trained in the frozen latent space.
Optimization uses Adam with learning-rate warmup and gradient-norm clipping,
and sampling uses an exponential moving average (EMA) of the denoiser
weights. The EDM noise schedule is shared by unconditional and conditional
sampling. The conditional-sampling guidance parameters and other training
and inference settings are summarized in Table~\ref{tab:supp_training}.

For unconditional generation, the conditioning context is set to zero,
classifier-free guidance is disabled by setting $w_{\mathrm{cfg}}=0$,
and no DPS correction is applied. The guidance weights reported in
Table~\ref{tab:supp_training} therefore correspond to conditional sampling.

% ------------------------------------------------------------
% Table S4: Training and sampling hyperparameters
% ------------------------------------------------------------

\begin{table}[htbp]
\centering
\small
\caption{
Training and sampling settings for the 2D and 3D models.
The upper block summarizes VAE training, the middle block summarizes
latent-diffusion and hard-data-conditioning training, and the lower block
lists the EDM noise schedule and conditional-sampling operating point.
Both 2D and 3D models use Heun's second-order solver with $N=32$ sampling
steps ~\cite{karras2022elucidating}. The parameter $\lambda_{\mathrm{cons}}$ denotes the weight of the
fine-resolution observation-consistency loss.
}
\label{tab:supp_training}

\begin{tabular}{lll}
\hline
Setting & 2D & 3D \\
\hline

\multicolumn{3}{l}{\textbf{VAE training}} \\
\hline

Optimizer / learning rate
& Adam / $10^{-4}$
& Adam / $10^{-4}$ \\

Batch size
& $32$
& $32$ \\

Epochs
& $1001$
& $600$ \\

Reconstruction loss
& Binary cross-entropy
& Tversky ($\alpha=0.3$, $\beta=0.7$) \\

KL weight $\beta_{\mathrm{KL}}$
& $10^{-6}$
& $10^{-4}$ \\

\hline
\multicolumn{3}{l}{\textbf{Latent diffusion / hard-data conditioning}} \\
\hline

Optimizer / learning rate
& Adam / $2\times10^{-4}$
& Adam / $2\times10^{-4}$ \\

Batch size
& $32$
& $32$ \\

Epochs
& $1001$
& $500$ \\

Learning-rate warmup
& $500$ steps
& $500$ steps \\

Gradient-norm clipping
& $1.0$
& $1.0$ \\

EMA decay
& $0.999$
& $0.999$ \\

CFG dropout $p_{\mathrm{uncond}}$
& $0.1$
& $0.1$ \\

Consistency weight $\lambda_{\mathrm{cons}}$
& $0.1$
& $0$ \\

Consistency loss
& Class-balanced BCE
& Class-balanced BCE
  ($\lambda_{\mathrm{cons}}=0$) \\

Wells per sample $N_w$
& $9$
& $1$--$2$ \\

\hline
\multicolumn{3}{l}{\textbf{EDM schedule and conditional-sampling settings}} \\
\hline

$P_{\mathrm{mean}},\,P_{\mathrm{std}}$
& $-1.2,\;1.2$
& $-1.2,\;1.2$ \\

$\sigma_{\min},\,\sigma_{\max}$
& $0.002,\;80$
& $0.002,\;80$ \\

$\rho$
& $7$
& $7$ \\

Sampling steps $N$ (Heun)
& $32$
& $32$ \\

Guidance $w_{\mathrm{cfg}},\,w_{\mathrm{dps}}$
& $3,\;4$
& $5,\;0.5$ \\

\hline
\end{tabular}
\end{table}

\bibliography{ref}

@article{jing2003review,
  title={A review of techniques, advances and outstanding issues in numerical modelling for rock mechanics and rock engineering},
  author={Jing, Lanru},
  journal={International Journal of Rock Mechanics and Mining Sciences},
  volume={40},
  number={3},
  pages={283--353},
  year={2003},
  publisher={Elsevier}
}

@article{lei2017use,
  title={The use of discrete fracture networks for modelling coupled geomechanical and hydrological behaviour of fractured rocks},
  author={Lei, Qinghua and Latham, John-Paul and Tsang, Chin-Fu},
  journal={Computers and Geotechnics},
  volume={85},
  pages={151--176},
  year={2017},
  publisher={Elsevier}
}

@article{lavoine2020discrete,
  title={A discrete fracture network model with stress-driven nucleation: Impact on clustering, connectivity, and topology},
  author={Lavoine, Etienne and Davy, Philippe and Darcel, Caroline and Munier, Raymond},
  journal={Frontiers in Physics},
  volume={8},
  pages={9},
  year={2020},
  publisher={Frontiers Media SA}
}

@article{hyman2014conforming,
  title={Conforming Delaunay triangulation of stochastically generated three dimensional discrete fracture networks: A feature rejection algorithm for meshing strategy},
  author={Hyman, Jeffrey D and Gable, Carl W and Painter, Scott L and Makedonska, Nataliia},
  journal={SIAM Journal on Scientific Computing},
  volume={36},
  number={4},
  pages={A1871--A1894},
  year={2014},
  publisher={SIAM}
}

@article{hyman2015dfnworks,
  title={dfnWorks: A discrete fracture network framework for modeling subsurface flow and transport},
  author={Hyman, Jeffrey D and Karra, Satish and Makedonska, Nataliia and Gable, Carl W and Painter, Scott L and Viswanathan, Hari S},
  journal={Computers \& Geosciences},
  volume={84},
  pages={10--19},
  year={2015},
  publisher={Elsevier}
}

@article{ho2022classifier,
  title={Classifier-free diffusion guidance},
  author={Ho, Jonathan and Salimans, Tim},
  journal={arXiv preprint arXiv:2207.12598},
  year={2022}
}

@article{wood1994simulation,
  title={Simulation of the von Mises Fisher distribution},
  author={Wood, Andrew TA},
  journal={Communications in statistics-simulation and computation},
  volume={23},
  number={1},
  pages={157--164},
  year={1994},
  publisher={Taylor \& Francis}
}

@article{chen2023fracture,
  title={Fracture network characterization with deep generative model based stochastic inversion},
  author={Chen, Guodong and Luo, Xin and Jiao, Jiu Jimmy and Jiang, Chuanyin},
  journal={Energy},
  volume={273},
  pages={127302},
  year={2023},
  publisher={Elsevier}
}

@article{lackner2003guide,
  title={A guide to CO2 sequestration},
  author={Lackner, Klaus S},
  journal={Science},
  volume={300},
  number={5626},
  pages={1677--1678},
  year={2003},
  publisher={American Association for the Advancement of Science}
}

@article{tsang2015hydrologic,
  title={Hydrologic issues associated with nuclear waste repositories},
  author={Tsang, Chin-Fu and Neretnieks, Ivars and Tsang, Yvonne},
  journal={Water Resources Research},
  volume={51},
  number={9},
  pages={6923--6972},
  year={2015},
  publisher={Wiley Online Library}
}

@article{karra2015effect,
  title={Effect of advective flow in fractures and matrix diffusion on natural gas production},
  author={Karra, Satish and Makedonska, Nataliia and Viswanathan, Hari S and Painter, Scott L and Hyman, Jeffrey D},
  journal={Water Resources Research},
  volume={51},
  number={10},
  pages={8646--8657},
  year={2015},
  publisher={Wiley Online Library}
}

@article{ferreira2022framework,
  title={A framework for upscaling and modelling fluid flow for discrete fractures using conditional generative adversarial networks},
  author={Ferreira, Carlos AS and Kadeethum, Teeratorn and Bouklas, Nikolaos and Nick, Hamidreza M},
  journal={Advances in Water Resources},
  volume={166},
  pages={104264},
  year={2022},
  publisher={Elsevier}
}

@article{ho2020denoising,
  title={Denoising diffusion probabilistic models},
  author={Ho, Jonathan and Jain, Ajay and Abbeel, Pieter},
  journal={Advances in neural information processing systems},
  volume={33},
  pages={6840--6851},
  year={2020}
}

@inproceedings{rombach2022high,
  title={High-resolution image synthesis with latent diffusion models},
  author={Rombach, Robin and Blattmann, Andreas and Lorenz, Dominik and Esser, Patrick and Ommer, Bj{\"o}rn},
  booktitle={Proceedings of the IEEE/CVF conference on computer vision and pattern recognition},
  pages={10684--10695},
  year={2022}
}

@article{deng2024characterization,
  title={Characterization of discrete fracture networks with deep-learning based hydrogeophysical inversion},
  author={Deng, Yaping and Kang, Xueyuan and Ma, Haichun and Qian, Jiazhong and Ma, Lei and Luo, Qiankun},
  journal={Journal of Hydrology},
  pages={130819},
  year={2024},
  publisher={Elsevier}
}

@article{berre2019flow,
  title={Flow in fractured porous media: A review of conceptual models and discretization approaches},
  author={Berre, Inga and Doster, Florian and Keilegavlen, Eirik},
  journal={Transport in Porous Media},
  volume={130},
  number={1},
  pages={215--236},
  year={2019},
  publisher={Springer}
}

@article{ringel2021stochastic,
  title={Stochastic inversion of three-dimensional discrete fracture network structure with hydraulic tomography},
  author={Ringel, Lisa Maria and Jalali, Mohammadreza and Bayer, Peter},
  journal={Water Resources Research},
  volume={57},
  number={12},
  pages={e2021WR030401},
  year={2021},
  publisher={Wiley Online Library}
}

@article{hu2022modeling,
  title={Modeling of solute transport in a fracture-matrix system with a three-dimensional discrete fracture network},
  author={Hu, Yingtao and Xu, Wenjie and Zhan, Liangtong and Zou, Liangchao and Chen, Yunmin},
  journal={Journal of Hydrology},
  volume={605},
  pages={127333},
  year={2022},
  publisher={Elsevier}
}

@book{zhang2002numerical,
  title={Numerical modelling and analysis of fluid flow and deformation of fractured rock masses},
  author={Zhang, Xing and Sanderson, David J},
  year={2002},
  publisher={Elsevier}
}

@article{berkowitz2002characterizing,
  title={Characterizing flow and transport in fractured geological media: A review},
  author={Berkowitz, Brian},
  journal={Advances in water resources},
  volume={25},
  number={8-12},
  pages={861--884},
  year={2002},
  publisher={Elsevier}
}

@book{bear2012flow,
  title={Flow and contaminant transport in fractured rock},
  author={Bear, Jacob and Tsang, Chin-Fu and De Marsily, Ghislain},
  year={2012},
  publisher={Academic Press}
}

@article{dhariwal2021diffusion,
  title={Diffusion models beat gans on image synthesis},
  author={Dhariwal, Prafulla and Nichol, Alexander},
  journal={Advances in neural information processing systems},
  volume={34},
  pages={8780--8794},
  year={2021}
}

@article{margolin1998structure,
  title={Structure, flow, and generalized conductivity scaling in fracture networks},
  author={Margolin, Gennady and Berkowitz, Brian and Scher, Harvey},
  journal={Water Resources Research},
  volume={34},
  number={9},
  pages={2103--2121},
  year={1998},
  publisher={Wiley Online Library}
}

@article{wang2020mrst,
  title={MRST-shale: an open-source framework for generic numerical modeling of unconventional shale and tight gas reservoirs},
  author={Wang, Bin},
  journal={Preprints},
  year={2020},
  publisher={MDPI AG}
}

@article{von2012lsd,
  title={LSD: A line segment detector},
  author={Von Gioi, Rafael Grompone and Jakubowicz, J{\'e}r{\'e}mie and Morel, Jean-Michel and Randall, Gregory},
  journal={Image Processing On Line},
  volume={2},
  pages={35--55},
  year={2012}
}

@inproceedings{karras2022elucidating,
  title={Elucidating the design space of diffusion-based generative models},
  author={Karras, Tero and Aittala, Miika and Aila, Timo and Laine, Samuli},
  booktitle={Advances in Neural Information Processing Systems (NeurIPS)},
  volume={35},
  pages={26565--26577},
  year={2022}
}

@inproceedings{chung2023diffusion,
  title={Diffusion posterior sampling for general noisy inverse problems},
  author={Chung, Hyungjin and Kim, Jeongsol and Mccann, Michael T and Klasky, Marc L and Ye, Jong Chul},
  booktitle={International Conference on Learning Representations (ICLR)},
  year={2023}
}

@book{deutsch1998gslib,
  title={{GSLIB}: Geostatistical Software Library and User's Guide},
  author={Deutsch, Clayton V. and Journel, Andr{\'e} G.},
  edition={2nd},
  year={1998},
  publisher={Oxford University Press},
  address={New York}
}

@article{strebelle2002conditional,
  title={Conditional simulation of complex geological structures using multiple-point statistics},
  author={Strebelle, Sebastien},
  journal={Mathematical Geology},
  volume={34},
  number={1},
  pages={1--21},
  year={2002},
  publisher={Springer}
}

@article{mariethoz2010direct,
  title={The Direct Sampling method to perform multiple-point geostatistical simulations},
  author={Mariethoz, Gr{\'e}goire and Renard, Philippe and Straubhaar, Julien},
  journal={Water Resources Research},
  volume={46},
  number={11},
  pages={W11536},
  year={2010},
  publisher={Wiley Online Library}
}

@article{miele2025diffusion,
  title={Diffusion models for multivariate subsurface generation and efficient probabilistic inversion},
  author={Miele, Roberto and Linde, Niklas},
  journal={Computers \& Geosciences},
  volume={207},
  pages={106076},
  year={2026},
  publisher={Elsevier},
  doi={10.1016/j.cageo.2025.106076}
}

@article{lee2025latent,
  title={Latent diffusion model for conditional reservoir facies generation},
  author={Lee, Daesoo and Ovanger, Oscar and Eidsvik, Jo and Aune, Erlend and Skauvold, Jacob and Hauge, Ragnar},
  journal={Computers \& Geosciences},
  volume={194},
  pages={105750},
  year={2025},
  publisher={Elsevier},
  doi={10.1016/j.cageo.2024.105750}
}

@article{ovanger2025statistical,
  title={A statistical study of latent diffusion models for geological facies modeling},
  author={Ovanger, Oscar and Lee, Daesoo and Eidsvik, Jo and Hauge, Ragnar and Skauvold, Jacob and Aune, Erlend},
  journal={Mathematical Geosciences},
  year={2025},
  publisher={Springer},
  doi={10.1007/s11004-025-10178-5}
}

@article{teng2025generating,
  title={Generating high-fidelity discrete fracture networks from low-dimensional latent spaces using generative adversarial network},
  author={Teng, Zheng and Wu, Hui and Zhang, Jize and Ju, Xin and Qi, Shengwen},
  journal={International Journal of Rock Mechanics and Mining Sciences},
  volume={196},
  pages={106301},
  year={2025},
  publisher={Elsevier},
  doi={10.1016/j.ijrmms.2025.106301}
}

@inproceedings{teng2024generative,
  title={A Generative Model of Discrete Fracture Networks Based on Latent Diffusion Model},
  author={Teng, Zheng and Wu, Hui and Zhang, Jize and Ju, Xin},
  booktitle={ARMA/DGS/SEG International Geomechanics Symposium},
  note={Paper ARMA-IGS-2024-0455},
  year={2024},
  organization={ARMA}
}

@article{teng2026projection,
  title={Projection-based low-dimensional parameterization method for three-dimensional discrete fracture networks through generative adversarial network},
  author={Teng, Zheng and Wu, Hui and Zhang, Kun and Zhang, Jize and Ju, Xin and Chen, Bin},
  journal={Journal of Rock Mechanics and Geotechnical Engineering},
  year={2026},
  note={In press},
  doi={10.1016/j.jrmge.2026.02.019},
  publisher={Elsevier}
}

@inproceedings{kingma2014autoencoding,
  title     = {Auto-Encoding Variational Bayes},
  author    = {Kingma, Diederik P. and Welling, Max},
  booktitle = {International Conference on Learning Representations},
  year      = {2014}
}

@inproceedings{goodfellow2014generative,
  title     = {Generative Adversarial Nets},
  author    = {Goodfellow, Ian J. and Pouget-Abadie, Jean and Mirza, Mehdi
               and Xu, Bing and Warde-Farley, David and Ozair, Sherjil
               and Courville, Aaron and Bengio, Yoshua},
  booktitle = {Advances in Neural Information Processing Systems},
  volume    = {27},
  pages     = {2672--2680},
  year      = {2014}
}

@article{laloy2018sgan,
  title   = {Training-Image Based Geostatistical Inversion Using a Spatial
             Generative Adversarial Neural Network},
  author  = {Laloy, Eric and H{\'e}rault, Romain and Jacques, Diederik
             and Linde, Niklas},
  journal = {Water Resources Research},
  volume  = {54},
  number  = {1},
  pages   = {381--406},
  year    = {2018},
  doi     = {10.1002/2017WR022148}
}

@article{difederico2025latent,
  title   = {Latent Diffusion Models for Parameterization of Facies-Based
             Geomodels and Their Use in Data Assimilation},
  author  = {Di Federico, Guido and Durlofsky, Louis J.},
  journal = {Computers \& Geosciences},
  volume  = {194},
  pages   = {105755},
  year    = {2025},
  doi     = {10.1016/j.cageo.2024.105755}
}

@article{wang2023prior,
  title   = {A Prior Regularized Full Waveform Inversion Using Generative
             Diffusion Models},
  author  = {Wang, Fu and Huang, Xinquan and Alkhalifah, Tariq A.},
  journal = {IEEE Transactions on Geoscience and Remote Sensing},
  volume  = {61},
  pages   = {1--11},
  year    = {2023},
  doi     = {10.1109/TGRS.2023.3337014}
}

@article{altman2014chemical,
  title={Chemical and hydrodynamic mechanisms for long-term geological carbon storage},
  author={Altman, Susan J and Aminzadeh, Behdad and Balhoff, Matthew T and Bennett, Philip C and Bryant, Steven L and Cardenas, M Bayani and Chaudhary, Kuldeep and Cygan, Randall T and Deng, Wen and Dewers, Thomas and others},
  journal={The Journal of Physical Chemistry C},
  volume={118},
  number={28},
  pages={15103--15113},
  year={2014},
  publisher={ACS Publications}
}

@article{yoon2017application,
  title={Application of a pore-scale reactive transport model to a natural analog for reaction-induced pore alterations},
  author={Yoon, Hongkyu and Major, Jonathan and Dewers, Thomas and Eichhubl, Peter},
  journal={Journal of Petroleum Science and Engineering},
  volume={155},
  pages={11--20},
  year={2017},
  publisher={Elsevier}
}

@article{yoon2008impact,
  title={Impact of nonaqueous phase liquid (NAPL) source zone architecture on mass removal mechanisms in strongly layered heterogeneous porous media during soil vapor extraction},
  author={Yoon, Hongkyu and Werth, Charles J and Valocchi, Albert J and Oostrom, Mart},
  journal={Journal of Contaminant Hydrology},
  volume={100},
  number={1-2},
  pages={58--71},
  year={2008},
  publisher={Elsevier}
}

@article{yoon2012highly,
  title={Highly parameterized inverse estimation of hydraulic conductivity and porosity in a three-dimensional, heterogeneous transport experiment},
  author={Yoon, Hongkyu and McKenna, Sean A},
  journal={Water Resources Research},
  volume={48},
  number={10},
  year={2012},
  publisher={Wiley Online Library}
}

@article{tavakoli2013comparison,
  title={Comparison of ensemble filtering algorithms and null-space Monte Carlo for parameter estimation and uncertainty quantification using CO2 sequestration data},
  author={Tavakoli, Reza and Yoon, Hongkyu and Delshad, Mojdeh and ElSheikh, Ahmed H and Wheeler, Mary F and Arnold, Bill W},
  journal={Water Resources Research},
  volume={49},
  number={12},
  pages={8108--8127},
  year={2013},
  publisher={Wiley Online Library}
}

@article{kim2021fast,
  title={Fast and scalable earth texture synthesis using spatially assembled generative adversarial neural networks},
  author={Kim, Sung Eun and Yoon, Hongkyu and Lee, Jonghyun},
  journal={Journal of contaminant hydrology},
  volume={243},
  pages={103867},
  year={2021},
  publisher={Elsevier}
}

@article{gostick2019porespy,
  title={PoreSpy: A python toolkit for quantitative analysis of porous media images},
  author={Gostick, Jeff T and Khan, Zohaib A and Tranter, Thomas G and Kok, Matthew DR and Agnaou, Mehrez and Sadeghi, Mohammadamin and Jervis, Rhodri},
  journal={Journal of Open Source Software},
  volume={4},
  number={37},
  pages={1296},
  year={2019}
}

@article{schnabel2007efficient,
  author  = {Schnabel, Ruwen and Wahl, Roland and Klein, Reinhard},
  title   = {Efficient RANSAC for Point-Cloud Shape Detection},
  journal = {Computer Graphics Forum},
  volume  = {26},
  number  = {2},
  pages   = {214--226},
  year    = {2007},
  doi     = {10.1111/j.1467-8659.2007.01016.x}
}

@inproceedings{salehi2017tversky,
  author    = {Salehi, Seyed Sadegh Mohseni and Erdogmus, Deniz
               and Gholipour, Ali},
  title     = {Tversky Loss Function for Image Segmentation Using
               3D Fully Convolutional Deep Networks},
  booktitle = {Machine Learning in Medical Imaging},
  series    = {Lecture Notes in Computer Science},
  volume    = {10541},
  pages     = {379--387},
  publisher = {Springer},
  year      = {2017},
  doi       = {10.1007/978-3-319-67389-9_44}
}
\bibliographystyle{abbrv}

\end{document}